\documentclass[twocolumn,showpacs,preprintnumbers,amsmath,amssymb,superscriptaddress]{revtex4}
\usepackage{graphicx}
\usepackage{dcolumn}
\usepackage{bm}
\usepackage{longtable}
\usepackage{subfigure}

\newcommand{\beq}{\begin{equation}}
\newcommand{\eeq}{\end{equation}}
\newcommand{\bea}{\begin{eqnarray}}
\newcommand{\eea}{\end{eqnarray}}

\begin{document}

\preprint{APS/123-QED}

\title{Organization and evolution of synthetic idiotypic networks}

\author{Elena Agliari}
\affiliation{Dipartimento di Fisica, Universit\`a degli Studi di
Parma, Italy}
\affiliation{INFN, Gruppo Collegato di
Parma}
\author{Lorenzo Asti}
\affiliation{Dipartimento di Scienze di Base e Applicate per l'Ingegneria, Sezione di Matematica, Sapienza Universit\`a di Roma,  Italy}
\affiliation{Dipartimento di Fisica, Sapienza Universit\`a di
Roma,  Italy}
\author{Adriano Barra}
\affiliation{Dipartimento di Fisica, Sapienza Universit\`a di
Roma,  Italy}
\author{Luca Ferrucci}
\affiliation{Dipartimento di Fisica, Sapienza Universit\`a di
Roma,  Italy}

\date{\today}

\begin{abstract}
We introduce a class of weighted graphs whose properties are meant to mimic the topological features of idiotypic networks, namely the interaction networks involving the B-core of the immune system. Each node is endowed with a bit-string representing the idiotypic specificity of the corresponding B cell and a proper distance between any couple of bit-strings provides the coupling strength between the two nodes.
We show that a biased distribution of the entries in bit-strings can yield fringes in the (weighted) degree distribution, small-worlds features, and scaling laws, in agreement with experimental findings. We also investigate the role of ageing, thought of as a progressive increase in the degree of bias in bit-strings, and we show that it can possibly induce mild percolation phenomena, which are investigated too.
\end{abstract}

 \maketitle

\section{Introduction}

Network theories are becoming fundamental pillars of modern approaches to describe reality in all fields of science, ranging from quantitative sociology \cite{granovetter,noi,callaway,pnas} to systemic biology \cite{reviews,libro,bio1,bio2}. As for the latter, immunology is one of the few fields of science where the importance of underlying networks of interactions were stressed directly by a few immunologists as early as half a century ago \cite{a8,jerne}:
The importance of refined models for the core of B-cell interactions is fundamental in order to properly address a possible regulatory role of the idiotypic network in "systemic" outcomes of the immune system, such as self-nonself discrimination \cite{a38,a39,perelson,kitamura}.
Furthermore, the recent goals obtained by monoclonal antibody techniques in clinical therapies \cite{monobook}  encourage the development of a clear theoretical backbone to control, for instance, the preparation and the effects of antibodies functionally mirroring proper enzymes \cite{luca1}
or the preparation of proper vaccinations against autoimmune manifestations (e.g. SLA) \cite{luca2}.

All this information continues to fascinate physicists and mathematicians who early started to develop theoretical models shaped to the increasing availability of read data (see e.g. \cite{perelsonbook,boer2}) and continue to update  plausible frameworks for these B-cell interactions (see e.g. \cite{brede,brete1,brete2,brasiliani} and references therein).

This paper presents a detailed study of the topological properties, and consequent structural implications, of a model for the B-cell branch of adaptive immunity, previously developed in \cite{BA,physa}.
Here, inspired by recent biological findings \cite{bialek2}, we introduce a tunable bias among antibodies at the epitopal (idiotypic) level, which aims to mimic the non purely random genesis of these proteins.
The inclusion of this feature has crucial consequences for the topology structure such as the emergence of fringes (i.e., multimodularity) in the weighted degree distribution; moreover, while with further increase in the correlation, the network gets sparser and sparser  (possibly reaching underpercolated regimes) \cite{luca4},  and proper scaling laws can be detected, all in agreement (at least qualitatively) with experimental results \cite{a39}.
In particular, fringes in degree distribution were observed in the first studies \cite{a39,stewart2} of antibody networks in mice and a structural role of these fringes was speculated: antibodies corresponding to nodes with a higher (weighted) coordination number would act as inhibitors to autoimmune errors (self-attaching), while those with a lower coordination number would act as soldiers against pathogens. Moreover, a relation between the overall similarity (i.e., the degree of bias) among antibody structures and the average coordination number has also been extensively reported \cite{luca4,stewart2,coutinho2,stewart3}, and it suggests the following speculation: the B-repertoire of an immune system of newborn mice is essentially random (so as to be highly adaptable to the outside reality), hence displaying a high connectivity. As infections are encountered, the network specializes its army, increasing antibody similarities \footnote{The simplest way to understand this statement is considering hyper-somatic mutation in lymphonodes during an immune response. There, antibodies able to attack strongly enough to the pathogen will undergo to somatic rearrangement to increase further their specificity against the enemy. As a consequence, as they will try to be the same "mirror" of the enemy, they will increase their correlation and decrease the reciprocal capability of binding.} and, as a consequence,  the network connectivity decreases, consistent with experimental findings in adult mice.

In this paper, after a streamlined introduction on the B-core of the immune system (Sec.~\ref{sec:introimmuno}), we introduce a class of weighted graphs meant to describe idiotypic networks (Sec.~\ref{sec:mod}). Then, such a model is investigated for its global topology (Sec. IV), including degree distribution and clustering features, and as for the distribution of weights associated witth its links (Sec. V).
Next the ageing of the system by increasing the degree of bias among bit strings; this process induces a progressive dilution of the network and the resulting percolation is analyzed (Sec.~\ref{sec:percola}).
Finally, we discuss our results and possible perspectives (Sec.~\ref{sec:concl}).

\section{A glance at adaptive immunity}\label{sec:introimmuno}
Mammalian immune system is a complex ensemble of interacting cells and exchanging molecules, whose scope is to prevent the host body from infection and damage by pathogens, either external (e.g. bacteria, fungi, viruses) or internal (e.g. cancerous cells). To satisfy this task, specific soldiers, each with specific weapons, were developed. The best known troop in this army is probably the family of B-cells, whose weapons are the antibodies (also known as immunoglobulin) they secrete.

In a nutshell, B-cells are divided into clones; each clone is made of cells producing the same antibody, and its size (the number of identical cells within the clone) varies over several orders of magnitude as a function of external stimulation. Consequently, the ensemble of all clones mirrors the repertoire of all possible antibodies, which, in order to cover a wide range of responses, must be enormously variegate [e.g. in humans it is $\mathcal{O}(10^{10})$]. In the following, each antibody is thought of as a bit string of fixed length, whose entries represent the presence (1) or absence (0) of the corresponding idiotope/paratope, i.e., the specific module devoted to the interaction and recognition of an epitope, a piece of an antigen. The latter is, by definition, any molecule able to trigger an immune response and can also be thought of as a bit-string of the same length \cite{BA}.

Now, following the pioneering investigation by Burnet \cite{a6,a7,a48} known as ``clonal selection theory'', (and assuming for the sake of clarity a binary representation in terms of bitstrings of information for both antigens and antibodies \cite{perelsonbook} that is deepened later), when an antigen (e.g. $01101100$) is introduced into the body, the B-cell expressing a high-matching antibody (e.g. $10010011$), after exposure to the antigen itself, starts the replicative process, increasing the size of the clone, and all its cells  secrete antibodies in a large quantity. The latter, being soluble molecules, rapidly explore the body deleting the pathogen. After the removal, the excess lymphocytes undergo apoptosis, shrinking the size of the clone to normal values \footnote{At this level of description we skip "details" as the development of memory cells or the isotopic switch of anticorpal subfamilies e.g. IgM,IgG,IgA.}.

One step forward, following the idiotypic network representation by Jerne \cite{jerne}, the antibody produced at a high concentration can itself be identified as a ``stranger" in the host body (as in the past it never appeared macroscopically, for otherwise, the pathogen could not affect the host already vaccinated) and trigger a response by another antibody with complementary entries to itself (the anti-antibody, e.g., $10011100$, namely, an external reproduction of the pathogen, but devoid of dangerous DNA/RNA), and so on, giving rise to the so-called "antibody cascade" commonly seen in experiments (see e.g., \cite{cazenave,coutinho1} and references therein). When an antibody is considered an antigen, the portion of the combinatorial site which is recognized by another antibody is called the idiotope.

Another step forward, approaching the Varela theory \cite{a38,a39}, is to allow the same antibody to be recognized by several clones,
namely, the idea of
``mirror of mirror'' is enlarged into an affinity matrix, in such a way that an antibody represented by a string, say $(0110110)$,
is reactive not only with its mirror $(1001001)$, but also with $(1001000)$, even
though to a smaller extent. This gives rise to a network whose nodes are B-cells and links connect clones secreting well-matching immunoglobulins.
This network has been shown to posses a regulatory role emerging from its non trivial topology and coupling distribution. Some of these features have already been evidenced in the context of unbiased antibody distributions, e.g. low-dose tolerance \cite{immuno}, generation of memory cells \cite{physa} and a unifying framework where the Burnet, Jerne, and Varela theories synergistically co-operate \cite{immuno,physa}. On the other hand, as we show here, some experimental evidences can be captured by biasing the bit strings representing the epitopal expression of antibodies. Hence, we address questions such as,
What is the underlying topology, if any? How does the topology affect the responses of the system? Does the network live above the percolation threshold, or is it made of several independent clusters? How do these quantities evolve with time? These are questions of crucial importance within a systemic approach to our understanding of immunological reality.

It should be remarked that, given the huge amount of antibodies making up the repertoire, and given the high intrinsic component of randomness in their genesis at the genetic level \cite{a11}, a pure random, unbiased formation of antibodies can be justified in a first approximation. Nonetheless, this approach underlies some limits which are now briefly summarized.
First, we notice that proteins (in a immunological context, antibodies) purely randomly generated would not fold (or at least not all) into stable and functional structures \cite{protein}.
Moreover, during the ontogenesis, each new B cell generated (and consequently each new antibody) is tested against self: in the bone marrow, newborn lymphocytes able to interact strongly with self molecules are deleted so to avoid autoimmune reactions \cite{a11}. As a consequence of this learning process, the final repertoire is biased.
A further bias is due to the continuous antigenic stimulation and consequent production and reshape of optimal antibodies. Therefore, antibody structures get biased both through their genomic origin and as a result of the several external stimuli (adaptation to evolution). The latter effect suggests that the "effective degree of bias" can be considered as a raw measure of aging in a mature system: as time elapses and new external antigens are experienced, the repertoire gets more and more specific and, accordingly, more and more biased; correspondingly the average connectivity is expected to decrease. In fact, for instance, in \cite{coutinho2}, the authors show that the high connectivity of newborn antibodies is lost in adult mitogen-reactive B-cell repertoires.

\section{The model}\label{sec:mod}
In a healthy human body at rest, it is estimated that the total
number of clones generated from a single B-cell (and therefore exhibiting the same idiotypicity) is about $10^2$ to $10^4$ and that the
total number of B-cells amounts to some $10^{12}-10^{14}$, hence, the distinct clones is about $10^{8}-10^{12}$. As for
antibodies, their number is about $10^{18}$ and the number of
idiotopes/paratopes \footnote{We stress that we are dealing with a symmetric theory and as in several other representations (e.g. \cite{perelsonbook} and reference therein) epitopes and paratopes are thought of as equivalent such that we will use indiscriminately the term idiotype for both of them.} from which they are built, determining the idiotypicity of a given antibody, is estimated to be relatively small, i.e., of the order of $10^2-10^3$.

For our purposes, immunoglobulins (Igs) and antigens can be described in the same way, namely, by means of a binary string $\xi$ encoding the particular sequence of idiotopes/paratopes that characterize their structure.
Therefore, by comparing the strings $\xi_i$ and $\xi_j$ of two given agents, labeled $i$ and $j$, respectively, one can determine their degree of affinity and state whether they are likely to recognize each other and, if so, how strong the interaction is.

Now, the specificity encoded by $\xi_i$ will characterize not only antibodies but also the clone, labeled $i$, secreting that kind of antibodies. In particular, our system is made up of an ensemble of $M \times N$ lymphocytes, which can occur in two states: \emph{quiescent}, i.e., secreting a low dose
of Ig; and \emph{firing}, i.e., secreting high doses of Ig.
The status of a given lymphocyte is therefore described by the binary variable
$\sigma_i^{\iota} = \pm 1$, where $i=1,...,N$ distinguishes the idiotipicity, while $\iota=1,...,M$, distinguishes lymphocytes belonging to the same
$i^{th}$ clone (in principle, $M$ may depend on the clone considered): $\sigma_i^{\iota} = +1$ corresponds to the firing state, while $\sigma_i^{\iota} = -1$ corresponds to the quiescent state.

In this way, if immunoglobulins $\xi_i$ and $\xi_j$ are affine enough, the corresponding clones $i$ and $j$ can be thought to be in ``interaction'', because  a firing state of clone $i$, yielding a high concentration of immunoglobulin $\xi_i$, will be detected by lymphocytes belonging to clone $j$, which will react properly. These facts can be envisaged by means of a network of mutually interacting lymphocytes.

Let us now introduce the main definitions of the model.
Antibodies and antigens are modeled as binary strings representing the
possible expression of $L$ idiotopes/paratopes. The assumption that each
antibody can be thought of as a string of the same length $L$ is based
on the  observation that ``all the gamma-globulins have structural
characteristic surprisingly similar'' \cite{a9}.
Hence, the $L$ idiotopes
\begin{eqnarray}
\nonumber
 \xi^1 = (1,0,0,...,0), \\
\nonumber
 \xi^2 = (0,1,0,...,0),  \\
 \nonumber
 ... \\
 \nonumber
 \xi^L = (0,0,0,...,1),
\end{eqnarray}
may act as a base in such a way that a generic antibody $\xi_i$ can be
decomposed as a linear combination
$\xi_{i}  = \lambda_{i}^{1} \xi^1+ \lambda_{i}^{2} \xi^2 + ... +
\lambda_{i}^{L} \xi^L$, with $\lambda_{i}^{\mu} \in (0,1)$ accounting
for the expression $(1)$ of the $\mu$th idiotope on
the $i$th antibody or its absense $(0)$.

We introduce biased strings by assuming that each entry $\mu$ of the $i^{th}$ string is
extracted randomly according to the  distribution
\begin{equation}\label{eq:p(xi)}
P(\xi_i^{\mu}=1) = \frac{(1+a)}{2}, \ P(\xi_i^{\mu}=0) = \frac{(1-a)}{2},
\end{equation}
with $a \in [-1,+1]$. In this way, when $a=0$, we recover the previous unbiased model \cite{BA} and, in general, the average similarity between a pair of strings can be tuned via $a$ as
$\langle \xi_i^{\mu} \xi_j^{\mu} \rangle \sim (1+a)^2/4$.
As we will see, Eq. (\ref{eq:p(xi)}) provides a basic way to bias the repertoire, which allows us to study the direct effects on the network performance; more refined models can of course be obtained.

Given a couple of clones, say $i$ and $j$, the
$\mu$th entries of the corresponding strings are said to be
complementary, iff $\xi_i^{\mu} \neq \xi_j^{\mu}$. Therefore, the
number of complementary entries $\chi_{ij} \in [0, L]$ can be written
as $\chi_{ij}=  \sum_{\mu = 1}^{L}
[\xi_i^{\mu} (1 - \xi_j^{\mu}) +\xi_j^{\mu} (1 - \xi_i^{\mu}) ]$.
Of course, $\chi_{ij}$ strongly depends on the correlation parameter $a$ and, in turn, it directly affects the affinity between $i$ and $j$. In fact,
the non-covalent forces acting among antibodies depend on the
geometry, on the charge distribution and on
hydrophilic-hydrophobic effects which give rise to an attractive
(repulsive) interaction for any complementary (non-complementary)
match. In principle, once the protein folding problem is solved \cite{protein}, the whole analysis of this kind of network could be extremely simplified by directly studying the VDJ genes and their reshuffling; however, as this bridge among micro and meso biological scenarios is lacking, we rely on "effective descriptions." In particular, we assume that each
complementary/non-complementary entry yields an attractive/repulsive contribution \cite{bagley,farmer}; the ratio between their intensities is denoted by the positive parameter $\alpha$. Hence, we
introduce the measure for the affinity between $\xi_i$ and $\xi_j$,
\begin{equation}
\label{eq:affinity}
f_{a,\alpha,L}(\xi_i,\xi_j|a) \equiv
 [\alpha \chi_{ij} - (L-\chi_{ij})],
 \end{equation}
which ranges from $-L$
(when $\xi_i = \xi_j$) to $\alpha L$ (when all entries are
complementary, i.e. $\xi_i^{\mu} \equiv 1- \xi_j^{\mu}, \, \mu=1,...,L$). Now, when the repulsive
contribution prevails, that is, $f_{a,\alpha,L} \leq 0$, the two antibodies do not see each other and
the coupling among the corresponding lymphocytes  $J_{ij} (a,\alpha,L)$ is set equal to 0; conversely, when $f_{a,\alpha,L} > 0$, we take as $J_{ij}$ the exponential of the affinity. This choice is the simplest trial able to mimic a key-lock mechanism for a sharp pattern recognition.
Thus, we have:
\bea \label{eq:J}
J_{ij} (a,\alpha,L)
&\equiv&  \Theta(f_{a,\alpha,L}(\xi_i,\xi_j|a)) \cdot
\nonumber\\
\phantom{A} &\phantom{=}&\cdot \exp[f_{a,\alpha,L}(\xi_i,\xi_j|a)],
\eea
where $\Theta(x)$ is the
discrete Heaviside function returning $1$, if $x>0$, and $0$, if $x
\leq 0$ \footnote{The effect of the theta function is to remove those links whose strength is mathematically different from zero, but, from a practical point of view are so weak that whatever level of noise would clean them as well.}.
Indeed, the expression in Eq.~(\ref{eq:J}) ensures that the coupling strength among lymphocytes spans several orders of magnitude ($J_{ij} \in [0, \exp(\alpha L)]$), as expected from experimental results \cite{carneiro}.
\newline
One could possibly introduce a proper normalization in order to fix an average value for the coupling strength, which in turn fixes a scale for the level of noise ruling the thermodynamics of the system. This procedure is allowed due to the fact that, as the system is a ferromagnet, the average coupling is positive definite and that, for a given size $N$, couplings display an upper bound. Nonetheless, the normalization is somehow arbitrary and does not qualitatively affect the behavior of the system. As for our current aims, we can neglect this and take Eq.~(\ref{eq:J}) as an effective definition for the coupling strength between node $i$ and node $j$.

We also stress that $N$ and $L$ are intrinsically connected to each other. This can be easily seen in the case where the match among antibodies had to be perfect for
reciprocal recognition; then, in order to reproduce all possible
antibodies obtained by the $L$ idiotopes/paratopes, the immune system would
need $N = 2^L$ lymphocytes. Here, having relaxed the hypothesis of perfect
match, only a fraction of this quantity needs to be retained to manage the whole
repertoire, and we can introduce the following scaling between
the number of all possible idiotypically different lymphocytes and the effective size of the repertoire:
\begin{equation}\label{NL}
L = \gamma \log N,
\end{equation}
where $\gamma >0$.

In order to quantify immune responses one can introduce the set of observables
\begin{equation}
m_i= \frac{1}{M}\sum_{\iota=1}^{M}\sigma_i^{\iota},
\end{equation}
where $i$
labels the clone, and $\iota$ labels the lymphocyte within each clone.
The thermodynamic of the system can then be described by means of an Hamiltonian,
\begin{equation}
\label{eq:H_1}
H = - \frac{1}{N} \sum_{i<j}^{N,N} J_{ij} m_i m_j - c \sum_i^N h^k_i m_i,
\end{equation}
where the first term on the right hand side represents the mutual interaction among lymphocytes, while the second term accounts for the interaction with an antigen $k$ present at concentration $c$ and whose coupling with the $i$-th clone is $h^k_i$ (see \cite{BA,aldo} for further details).
Interestingly, these two terms encode the Jerne and Burnet theories, respectively.

\section{Global topology}\label{sec:global}
If we forget, for the moment, the weights of the couplings, the $N$ different B clones, interacting pairwise, define a graph $\mathcal{G}= \{V,\Gamma \}$, where $V$ denotes the set of nodes and $\Gamma$ the set of links. The cardinality of $V$ is given by $|V| = N$, that is the total amount of idiotypically different clones. The topological properties of $\mathcal{G}$
are completely determined by the adjacency matrix $A$ defined as $A_{ij}=1$ if $J_{ij}\neq 0$ and $A_{ij}=0$ if $J_{ij}=0$. For instance, the degree of a node $i$ (i.e.  coordination number) is given by $z_{i} = \sum_{j\in V}A_{ij}$.
In the following, we provide the main definitions and formula to describe the topology of the emergent graph and later we will deepen its global features.

\begin{figure}[tb]\label{fig:P_k}
\resizebox{0.485\columnwidth}{!}{\includegraphics{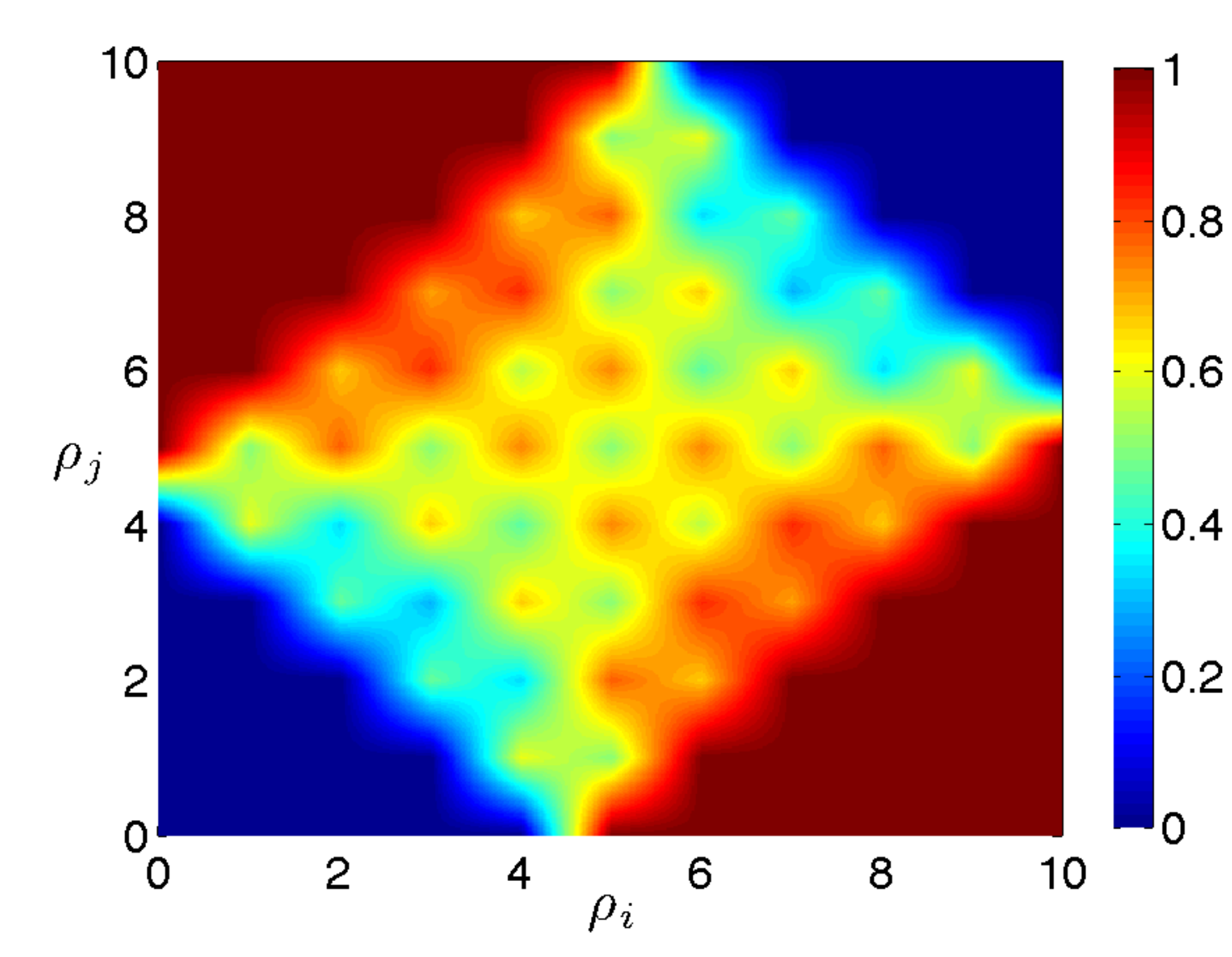}}
\resizebox{0.495\columnwidth}{!}{\includegraphics{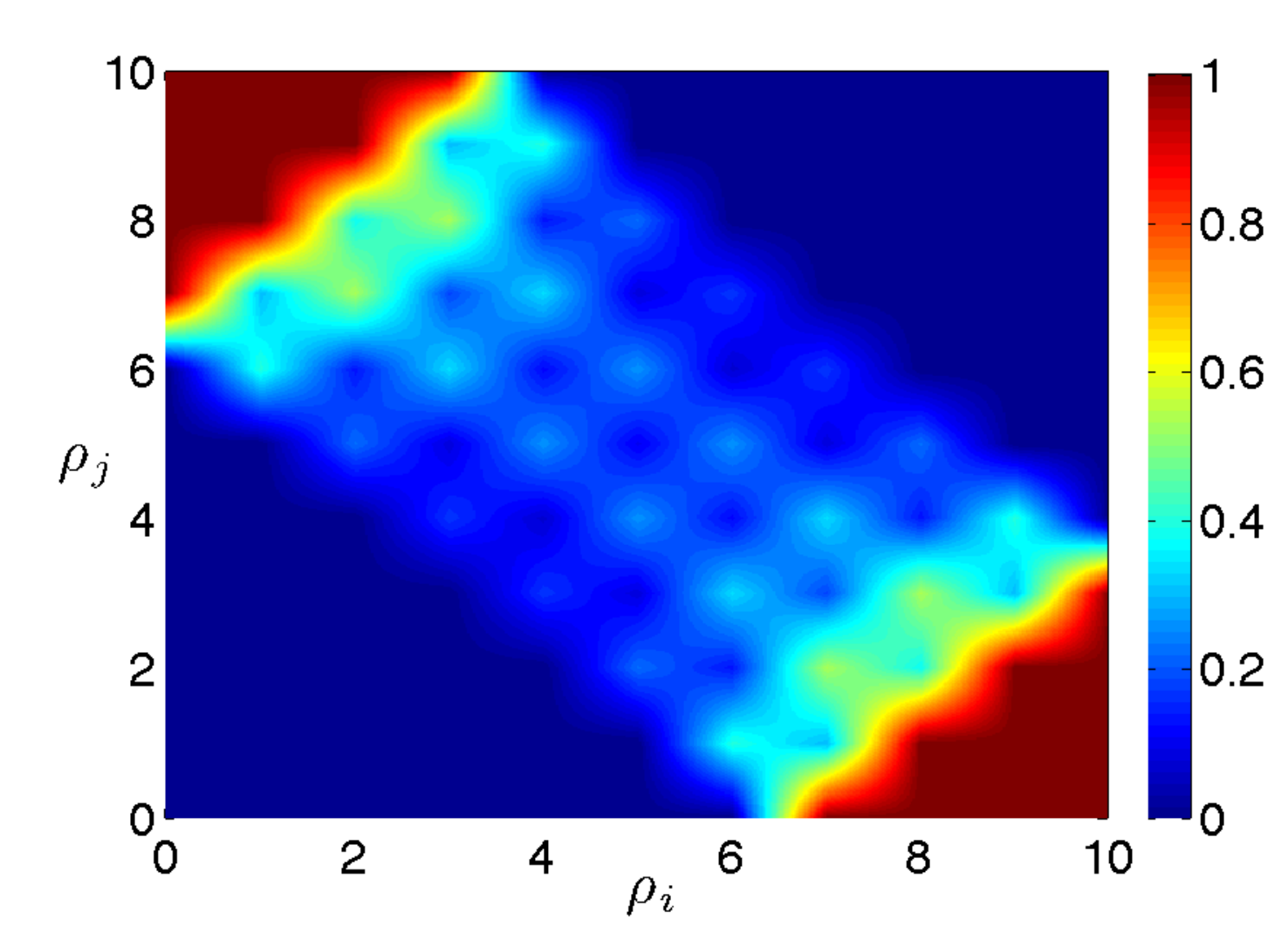}}
\caption{(Color on line) Link probability $P_{\textrm{link}}(\rho_i, \rho_j; \alpha, L)$ as a function of $\rho_i$ and $\rho_j$ with $L=10$ and $\alpha = 0.5$ (left) or $\alpha = 1.1$ (right).}
\label{fig:p_roiroj}
\end{figure}

First, let us introduce the
probability that a string $\xi_{i}$ displays $\rho$ non null entries; this follows a binomial distribution
\begin{equation}\label{eqn:prob_ro}
P(\rho;a,L) = {L \choose \rho} \left( \frac{1+a}{2} \right)^\rho \left(\frac{1-a}{2} \right)^{L-\rho}.
\end{equation}
Correspondingly, the probability that two strings $\xi_{i}$ and $\xi_{j}$, displaying $\rho_i$ and $\rho_j$ non-null entries respectively, exhibit $\chi$ complementary matches is
\begin{eqnarray}\label{eqn:prob_accoppiamento}
&&P(\chi; \rho_{i},\rho_{j}) = \binom{L} {\rho_i}^{-1} \binom{L}{ \rho_j}^{-1} \times \\
\nonumber
&&\frac {L!}{ \left ( \frac{\rho_i - \rho_j + \chi}{2} \right)! \left( \frac{\rho_j - \rho_i + \chi }{2} \right)! \left ( \frac{\rho_i+ \rho_j - \chi }{2}\right)! \left(L-\frac{\rho_i + \rho_j + \chi}{2} \right)! }.
\end{eqnarray}
Now, the link probability $P_{\textrm{link}}( \rho_{i}, \rho_{j}; \alpha, L)$ can be obtained by summing $P(\chi; \rho_{i},\rho_{j})$ over the values of $\chi$ compatible with $(\rho_{i}, \rho_{j})$ and such that  $f_{\alpha,L}(\xi_i, \xi_j)>0$, namely, recalling Eq.~(\ref{eqn:prob_accoppiamento}),
\begin{equation}\label{eqn:prob_link}
P_{\textrm{link}}(\rho_{i},\rho_{j};\alpha,L) = \sum_{\chi = \max (| \rho_{i}-\rho_{j}|, \ \lceil \frac{L}{\alpha+1}\rceil)}^{\min(\rho_{i}+\rho_{j}, \ 2L - \rho_{i} - \rho_{j})} \!\!\!P(\chi;\rho_{i},\rho_{j})
\:.
\end{equation}
Figure \ref{fig:p_roiroj} shows an example of $P_{\textrm{link}}(\rho_{i},\rho_{j};\alpha,L)$ for different choices of the parameter $\alpha$.

By averaging $P_{\textrm{link}}(\rho_{i},\rho_{j};\alpha,L)$  over $\rho_i$, we get the average link probability for a node whose string displays $\rho_{i}$ non-null entries, that is,
\begin{equation}\label{eq:P_roi}
P_{\textrm{link}}(\rho_{i};a, \alpha,L)=  \sum_{\rho_{j}=0}^{L}P(\rho_{j};a,L)P_{\textrm{link}}(\rho_{i},\rho_{j};\alpha,L),
\end{equation}
from which the average degree for $j$ reads
\begin{equation}
z(\rho_{i};a, \alpha,L,N)= N P_{\textrm{link}}(\rho_{i};a, \alpha,L).
\end{equation}
Numerical calculations of $P_{\textrm{link}}(\rho; a,\alpha, L)$ are shown in Fig.~$2$: Notice that a uniform bit distribution within the antibodies (i.e. $a=0$) corresponds to an unbiased graph, where the average link probability of a node does not depend on the pertaining string.
\newline
By further averaging $P_{\textrm{link}}(\rho_{i};a, \alpha,L)$ over $\rho_{i}$, we get the average link probability for an arbitrary node of the system
\begin{eqnarray}\label{eqn:prob_a_alpha}
\nonumber
\overline{P}_{\textrm{link}}(a,\alpha,L) &=& \sum_{\rho_i,\rho_j=0}^{L,L}P_{\textrm{link}}(\rho_{i},\rho_{j};\alpha,L)  \times\\
&\phantom{=}& \times \, P(\rho_i;a,L)P(\rho_j;a,L),
\end{eqnarray}
from which the average coordination number follows as
\begin{equation}\label{eq:average_degree}
\overline{z}(a, \alpha,L,N)= N \overline{P}_{\textrm{link}}(a, \alpha,L).
\end{equation}
%
\begin{figure}[!]\label{fig:P_ro}
\resizebox{0.493\columnwidth}{!}{\includegraphics{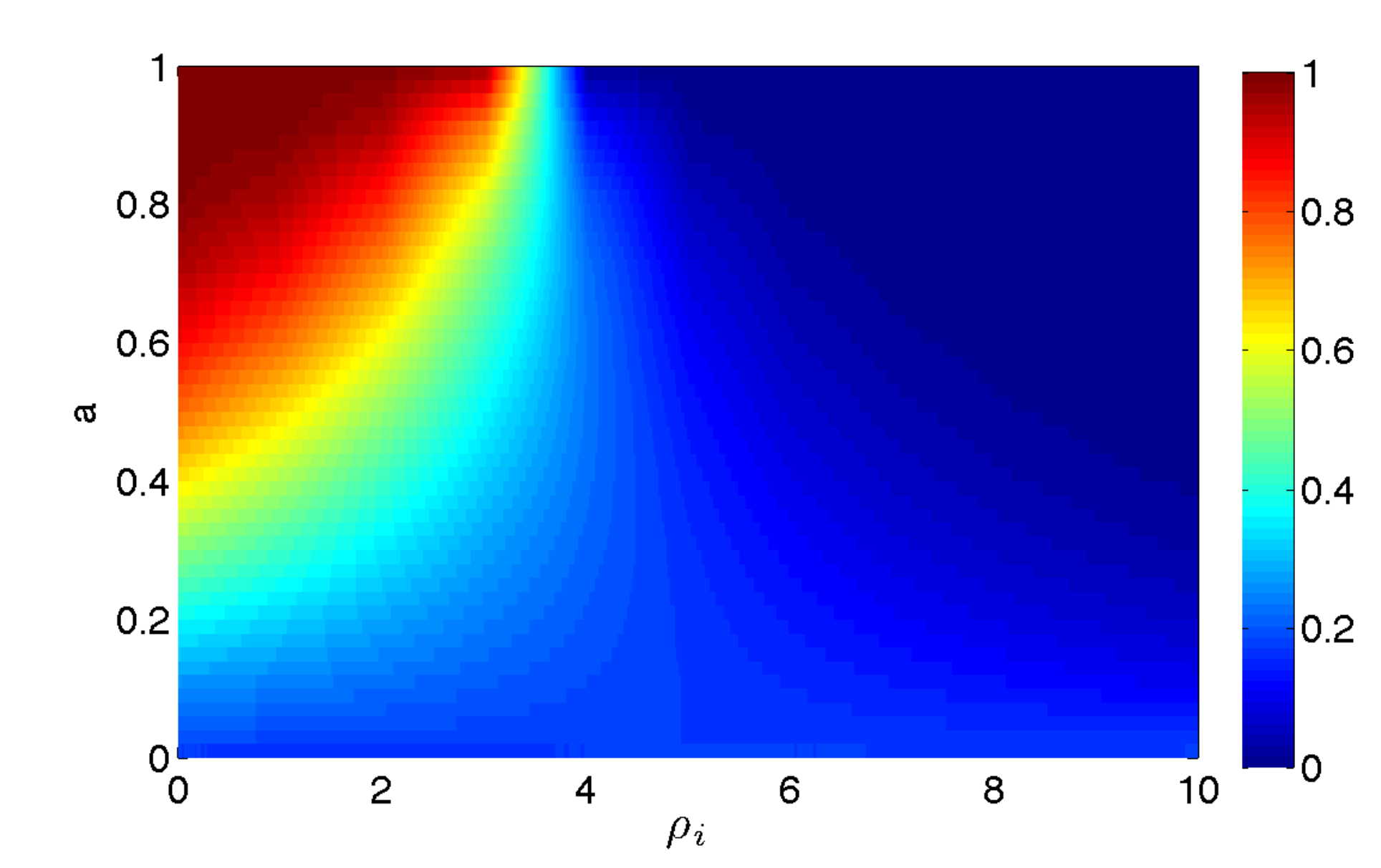}}
\resizebox{0.493\columnwidth}{!}{\includegraphics{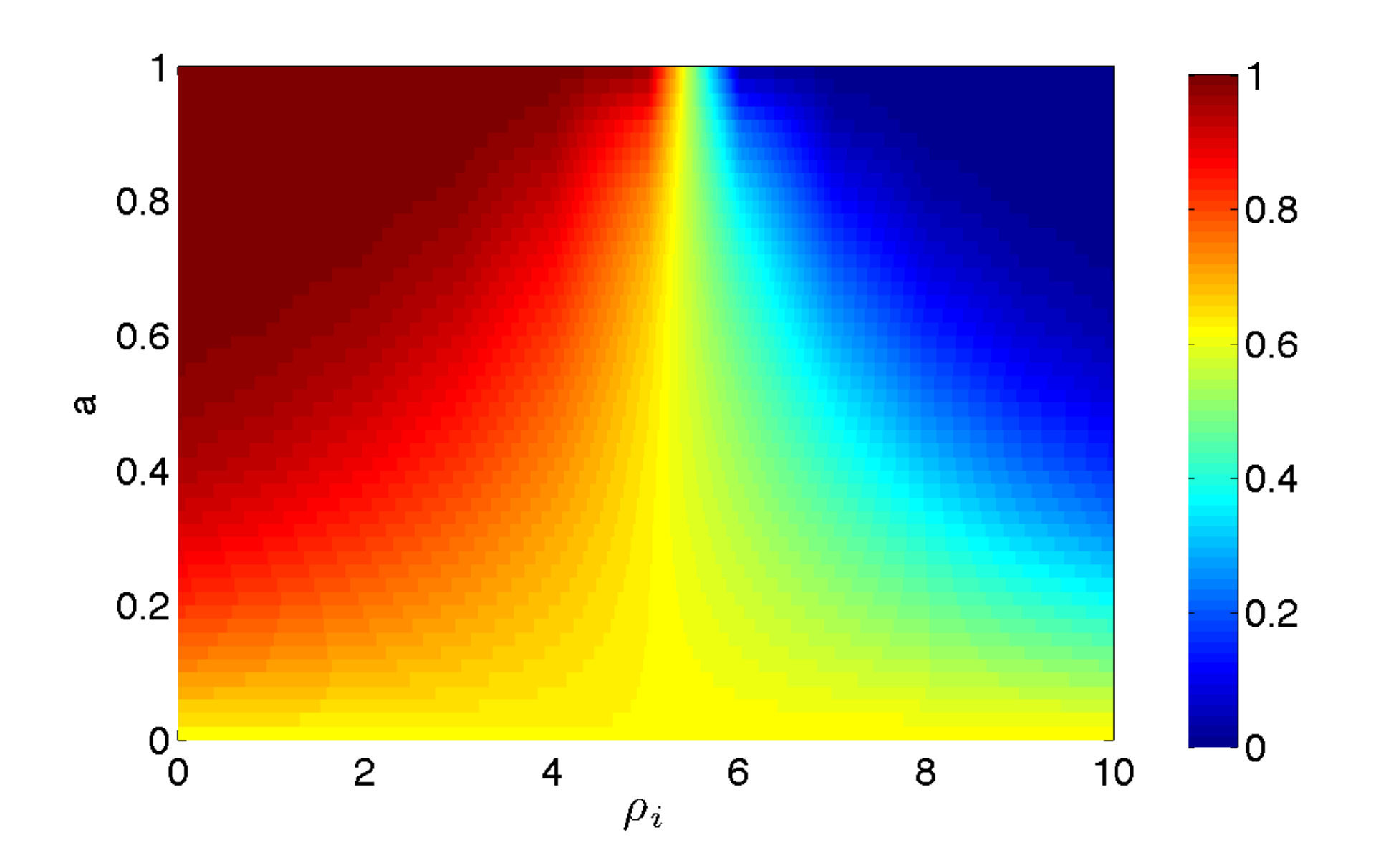}}
\caption{(Color on line) Link probability $P_{\textrm{link}}(\rho_i; a, \alpha, L)$  as a function of $\rho_{i}$ and $a$, while $\alpha$ and $L$ are fixed. In the left panel $\alpha=0.5$ while in the right panel $\alpha=1.1$; in both cases $L=10$. It is straightforward to see that on the line $a=0$ the probability link is independent on $\rho_{i}$. When $a \neq 0$, the link probability changes with varying $\rho_{i}$. Due to symmetry, only the range $a \in [0,1]$ is shown.}
\end{figure}
Finally, within a mean-field approach, we can use $P_{\textrm{link}}(\rho;a,\alpha,L)$ to write the degree distribution:
\begin{eqnarray}\label{eq:P_degree_z}
\nonumber
&&P_{\textrm{degree}}(z|\rho;a,\alpha,L,N) = {N-1 \choose z} \left[P_{\textrm{link}}(\rho;a, \alpha,L)  \right]^{z} \\
&\times& \left[1-P_{\textrm{link}}(\rho;a,\alpha,L) \right]^{N-1-z},
\end{eqnarray}
and, by further averaging over $\rho$,
\begin{eqnarray}\label{eq:prog_degree}
\overline{P}_{\textrm{degree}}(z;a,\alpha,L,N) &=& \sum_{\rho=0}^{L} P_{\textrm{degree}}(z|\rho;a,\alpha,L,N)
\times
\nonumber\\
\phantom{A} &\phantom{=}& \phantom{\sum}\times P(\rho;a,L).
\end{eqnarray}
%

\subsection{Multimodal degree distribution}\label{sec:multimodal}
As shown by Eq.~(\ref{eq:prog_degree}), $\overline{P}_{\textrm{degree}}(z;a,\alpha,L,N)$ is the sum of $L$ binomial distributions, each referring to a different ``mode'' $\rho$.
Therefore, the average degree distribution will show a multimodal behavior as long as two consecutive modes have disjoint supports.

\begin{figure*}[!tbp]
\centering
{\includegraphics[scale=0.6]{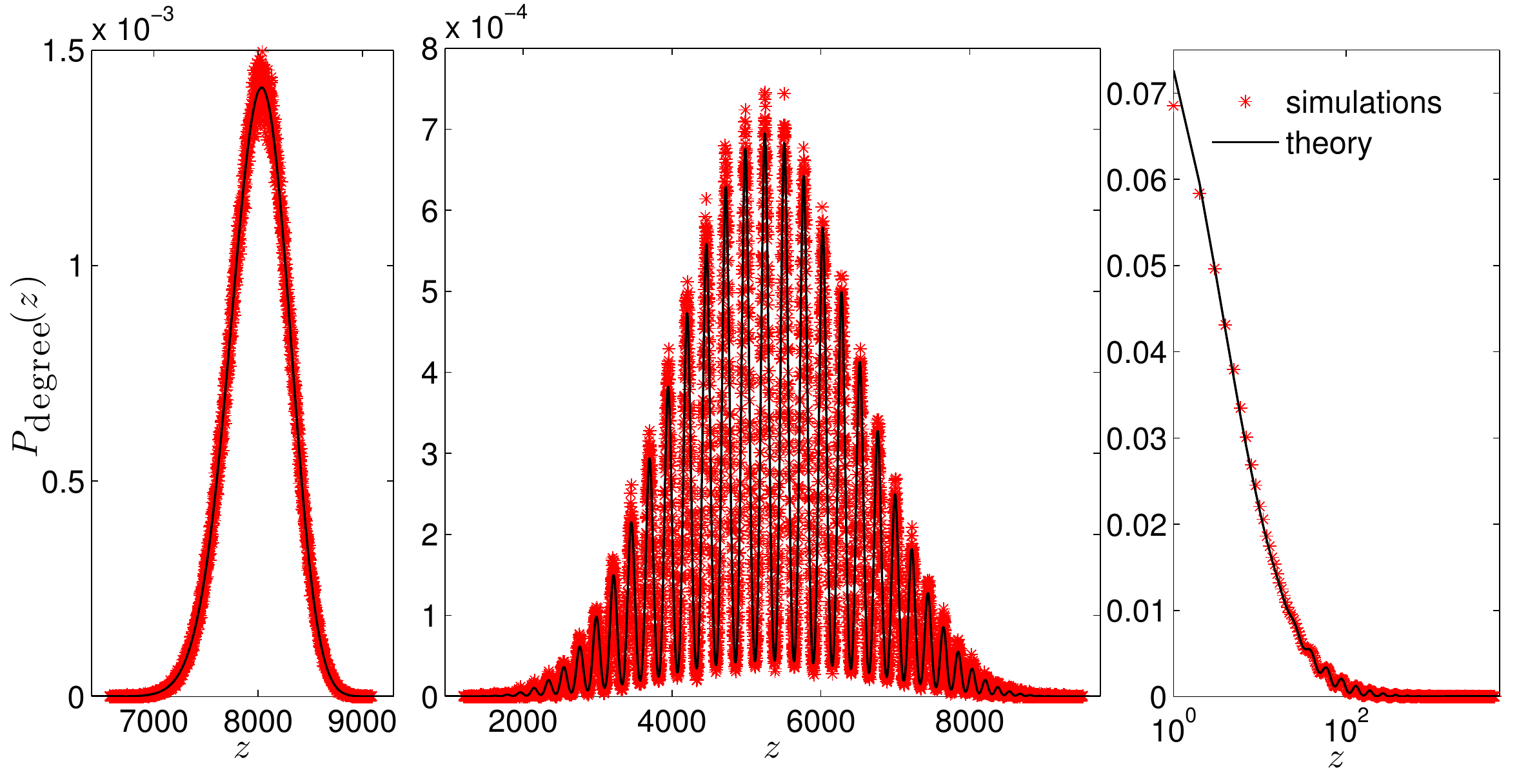}}
\caption{(Color on line) Degree distribution for different values of $a$ and fixed $\alpha=1.2$, $N=10000$, $\gamma=10$. From left to right: $a=0.1$ (uni-modal behavior; the mean degree is about $0.8 N$), $a=0.3$ (multimodal behavior; the mean degree is $0.5N$), and $a=0.6$ (there is no extensive network; the average degree is $0.03N$).  \label{fig:degree_N10000} }
\end{figure*}

In general we expect that, when $a$ is close to 0, peaks merge since, at low bias, the link probability weakly depends on $\rho$ (see also FIg. \ref{fig:P_ro}), which is uniformity, while, when $a$ is close to $1$, most of the modes are off and the network is sparse. Therefore a very multimodal distribution is expected only for intermediate values of $a$. Indeed, these arguments are corroborated by Fig.~(\ref{fig:degree_N10000}) which shows different plots of $\overline{P}_{\textrm{degree}}(z;a,\alpha,L,N)$ for different values of $a$; also notice that numerical data are finely fitted by the analytical calculation of Eq.~(\ref{eq:prog_degree}).

Consistent with the remarks in Sec.~I, an increase in $a$ basically encodes a set of ageing effects;  in fact we can envisage that, in the antenatal period, when lymphocytes can not encounter antigens, $\emph{a}$ is close to $0$, and the immune system is not specialized in any particular antigen response: There is no multi-modal behavior in the degree distribution and the underlying B-cell network is roughly uncorrelated.
Then, after the birth, lymphocytes start matching some antigens and undergo clonal expansion and, consequently, hypersomatic mutations: the system begins to specialize, and the connectivity starts to decrease. The degree distribution starts to exhibit a fine structure, where peaks correspond to specific sub-populations that have survived as a consequence of the antigen encountered in the past. 

\subsection{Scaling in thermodynamics limit. \label{sec:scal}}

As evidenced in the previous subsection, $a$ crucially affects the topology of the idiotypic network. We therefore investigate in more detail the global connectivity of the system in terms of $\overline{P}_{\textrm{link}}(a,\alpha,L,N)$. To compute this quantity, one would plug Eq.~(\ref{eqn:prob_accoppiamento}) into Eq. (\ref{eqn:prob_link}) and (\ref{eqn:prob_a_alpha}). However, since here we are interested in the large $L$ limit and relative fluctuations in $\rho$ decrease as $1/\sqrt{L}$, we can adopt a mean-field like approach and approximate Eq. (\ref{eqn:prob_ro}) as
\beq\label{eqn:rhomedio}
P(\rho; a, L) \:\simeq\: P_{MF}(\rho; a, L) = \delta(\rho - \bar{\rho}),
\eeq
with $\bar{\rho}=(1+a)L/2$, so that Eq.~(\ref{eqn:prob_accoppiamento}) can be restated as
\beq\label{pMF(chi)}
P_{MF}(\chi;  a, L)=
\frac{ \binom{(1+a)L/2}{\chi/2} \binom{(1-a)L/2}{\chi/2} }{\binom{L}{(1+a)L/2}}.
\eeq
Moreover, exploiting the parity symmetry for $a$, we focus on the range $a \in [0,1]$.
Thus, Eq. (\ref{eqn:prob_link}) can be approximated as
\bea\label{eqn:p(chi)MF}
\!\!\overline{P}_{\textrm{link}}(a,\alpha,L,N) &\approx&
\nonumber\\
\!\!\overline{P}_{\textrm{link}}^{MF}(a,\alpha,L,N) &=&
\frac{ \sum_{\chi=\lceil \frac{L}{2(\alpha+1)} \rceil}^{(1-a)L} {\frac{1+a}{2}L \choose \chi/2}{\frac{1-a}{2}L  \choose \chi/2}}{ {L \choose \frac{1+a}{2}L  }}.
\eea
Now, since we are interested in scaling laws, we can neglect all terms in the sum, but the leading one. For instance, focusing on $a<1/2$ ($a>-1/2$) and $\alpha >1$, it is easy to see that this is given by $\chi=L/2$. Then, via Stirling approximation, we get
\begin{equation}\label{eq:f(a)_g(a)}
\log \overline{P}_{\textrm{link}}(a,\alpha,L,N)  \sim f(a)L,
\end{equation}
where $f(a) = (1+a)\log (1+a) +(1-a) \log (1-a) -(1+2a)/4 \log (1+2a) - (1-2a)/4 \log (1-2a)$ is a symmetrical monotonically decreasing function from $a=0$ to $a=\pm 1/2$ and roughly plays the role of an (compressed in the interval $[-1/2,+1/2]$) entropy of the bit-strings, such that, as in other biased approaches \cite{amit_low_level_activity}, $a$ can be thought of as the "bit-string magnetization".

Therefore, we can write
\begin{equation}\label{eq:f(a)}\overline{P}_{\textrm{link}}(a,\alpha,L,N)  \sim N^{\gamma f(a)}.
\end{equation}
In this way, as the degree of bias $a$ is increased, the graph turns from highly connected ($\overline{P}_{\textrm{link}} = \mathcal{O}(1)$) to diluted ($\overline{P}_{\textrm{link}} \sim N^{-\gamma c}$, with $c=\log(27/32)/2$).
Similar results can be found for a different parameter range.

\subsection{Clustering Coefficient}
The local clustering coefficient is defined as the number of triangles stemming from a node \emph{i} over the maximum number of triplets centered on \emph{i} itself (see, e.g., Refs. \cite{ACG,newman,callaway}). Due to the ``anti-transitive'' nature of the idyotipic network, triangles, i.e., $3$-cycles, are expected to be unlikely, while quadrilaterals, i.e. $4$-cycles, are expected to be favored. Roughly speaking, an antibody Ab1 elicits its anti-antibody Ab2, which in turn elicits the anti-anti-antibody Ab3, whose structure should be close to Ab1's, so that a $4$-cycle finally develops.

\begin{figure}[!]
\centering
\resizebox{0.9\columnwidth}{!}{\includegraphics{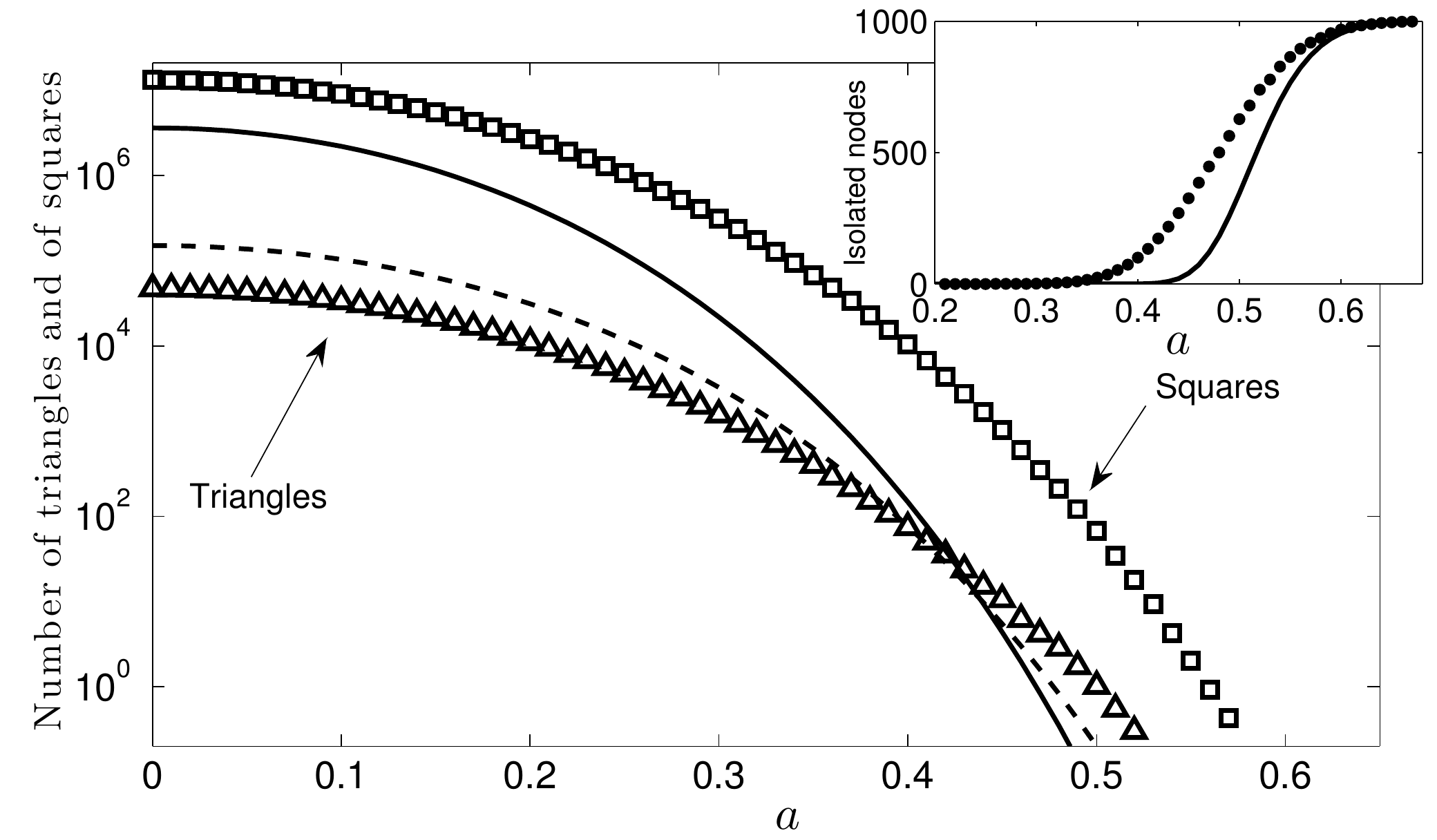}}
\caption{Number of triangles ($\triangle$) and of quadrilaterals ($\square$) averaged over $100$ realizations in our idiotypic network as a function of $a$. Parameters characterizing the network are $N=1000$, $\gamma=7$ and $\alpha=0.7$. Curves represent the number of triangles (dashed line) and of squares (solid line) expected for an analogous ER graph. Inset: Number of isolated nodes present in the system; again, the idiotypic network ($\bullet$) and ER graph (line) are compared. The latter displays a larger number of triangles as long as a giant component can be detected.}
\label{fig:motifs}
\end{figure}

Indeed, this was partially shown in Ref. \cite{BA}, where an idiotypic network and an analogous Erd\"{o}s-R\'{e}ny (ER) graph, namely, a purely random graph exhibiting the same average degree $\bar{z}$ were compared, obtaining that the former displays on average a significantly smaller number of triangles.
We now extend those results considering also $4$-cycles (devoid of diagonals).
Given a node $i$ with $z_{i}$ nearest neighbors, the number of expected squares stemming from $i$, in the case where $i$ belongs to a random ER graph and in the case where it belongs to our idiotypic network, are respectively
\begin{eqnarray}
Q_{ER}(z_i)=\binom{z_{i}}{2} (N-1-z_{i}) p^2 (1-p), \\
Q(z_i)=\binom{z_{i}}{2} (N-1-z_{i}) p'^2 (1-p''),
\end{eqnarray}
where $p=\bar{z}/N$, $p'$ is the probability that in $\mathcal{G}$ a neighbor of $i$ is linked to a node not belonging to the $i$th neighborhood, and $p''$ is the probability that two neighbors are linked.
The latter is just the clustering coefficient for node $i$, which, as shown in Ref. \cite{BA}, is smaller for a graph where links are based on complementarity features, so that $(1-p) \leq (1-p'')$.
Moreover, the idiotypic and the ER graphs we are comparing display, by construction, the same coordination number. If we impose this condition to be true also for the average degree of any site $j$ that is linked with $i$, we get
\begin{eqnarray}\label{eq:kmean}
\bar{z_j} &=& 1 + p^{''}(z_{i} - 1) + p{'}(N-z_{i}-2) \\
\nonumber
&=& 1 + p(z_{i} - 1) + p(N-z_{i}-2).
\end{eqnarray}
Therefore, as $p > p''$, we get $p < p'$ and finally $Q(z_i) > Q_{ER}(z_i)$. More generally, this suggests that in our idiotypic networks, as links are based on complementarity, $4$-cycles are motifs while $3$-cycles are anti-motifs \cite{BA,AB}.
Indeed, Fig.~\ref{fig:motifs} numerically confirms that the number of quadrilaterals [triangles] appearing in our graph is larger [smaller] than the number expected for an analogous ER graph, estimated as $\binom{N}{4}p^4 (1-p)^2$, $[\binom{N}{3}p^3]$.

\section{Coupling and weighted degree distributions}
In Sec. \ref{sec:global} we studied the bare topology of the network, while here we focus on the properties related to the distribution of weights $J_{ij}$ associated with links $(i,j)$. As mentioned above, these features retain a strong biological meaning. For instance, nodes displaying a high weighted degree feel, under normal conditions, a larger (internal) quiescent stimulus \cite{immuno}.

\subsection{Coupling distribution}\label{sec:couplings}
Given two nodes $i$ and $j$ with $\rho_i$ and $\rho_j$ non-null entries, respectively, recalling Eqs.~($2$) and ($3$), the coupling distribution is
\bea\label{eq:p(J)}
\nonumber
&&\!\!\!\!P(J_{ij}| \rho_{i}, \rho_{j}; \alpha,L) =
\nonumber\\
&&\!\!\!\! = \left \{
\begin{array}{cl}
P(\chi_{ij} = \frac{\log(J_{ij}) + L}{\alpha +1};\rho_i,\rho_j,L) & \mbox{if } J_{ij} > 1 \\
0 & \mbox{if } J_{ij} \leq 1
\end{array}
\right.
\,,
\eea
where $J_{ij}$ can span over $[1,e^{\alpha L}]$.

\begin{figure}[!]
\centering
\resizebox{1.0\columnwidth}{!}{\includegraphics{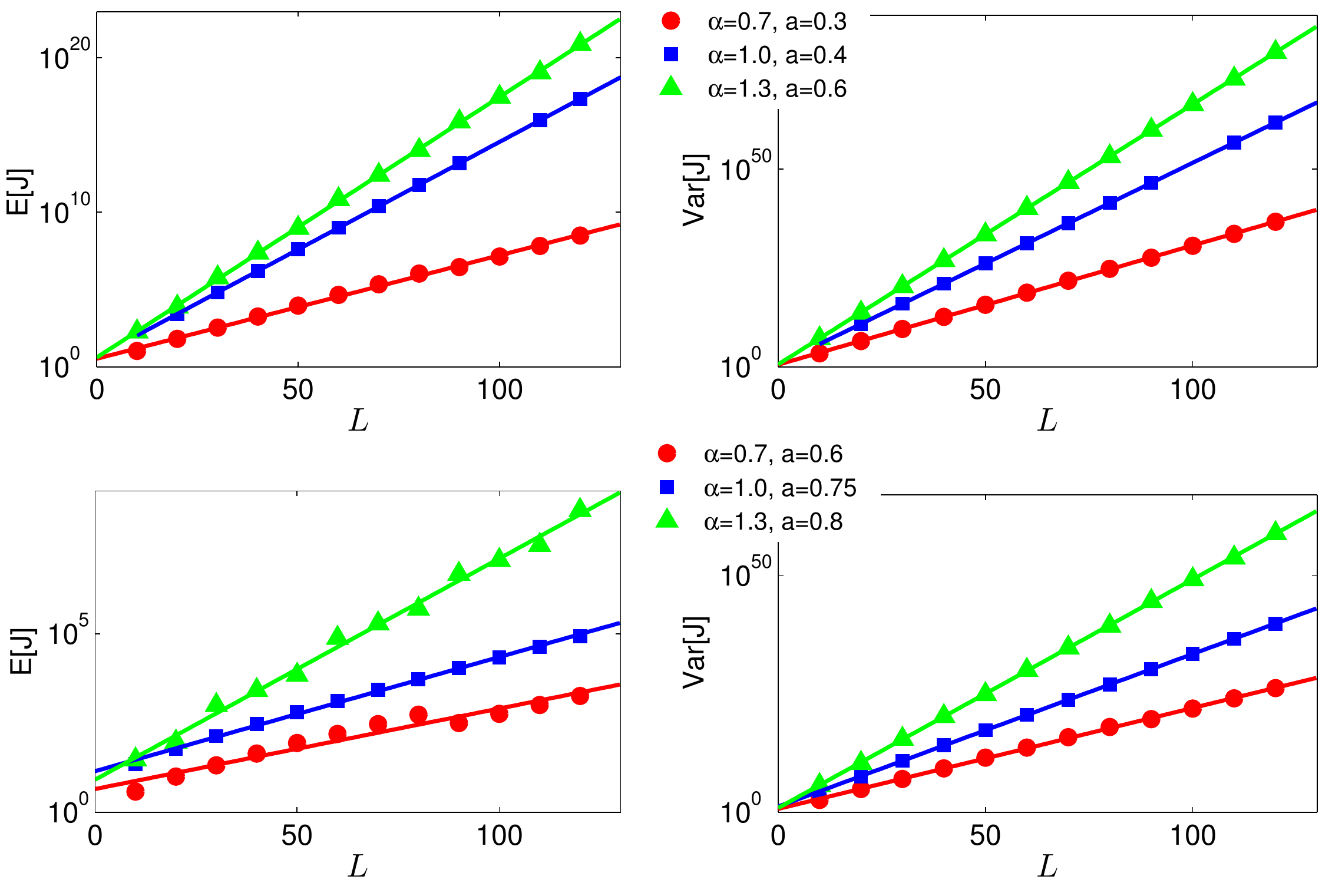}}
\caption{(Color on-line) Average value of the coupling strength E[J] and of its variance Var[J], versus $L$ for several choices of parameters, depicted in different colors. We considered  $\alpha=0.7$ ($\circ$), $\alpha=1.0$ ($\square$) and $\alpha=1.3$ ($\triangle$) for different values of $a$ pertaining to a connected (upper panels) and disconnected (lower panels) regimes (see the legend). Symbols refer to data obtained from exact numerical calculations, while curves are drawn according to the approximation~\ref{eq:J_scaling}.
}
\label{fig:J_vs_L}
\end{figure}

\begin{figure}[!]
\centering
\resizebox{1.0\columnwidth}{!}{\includegraphics{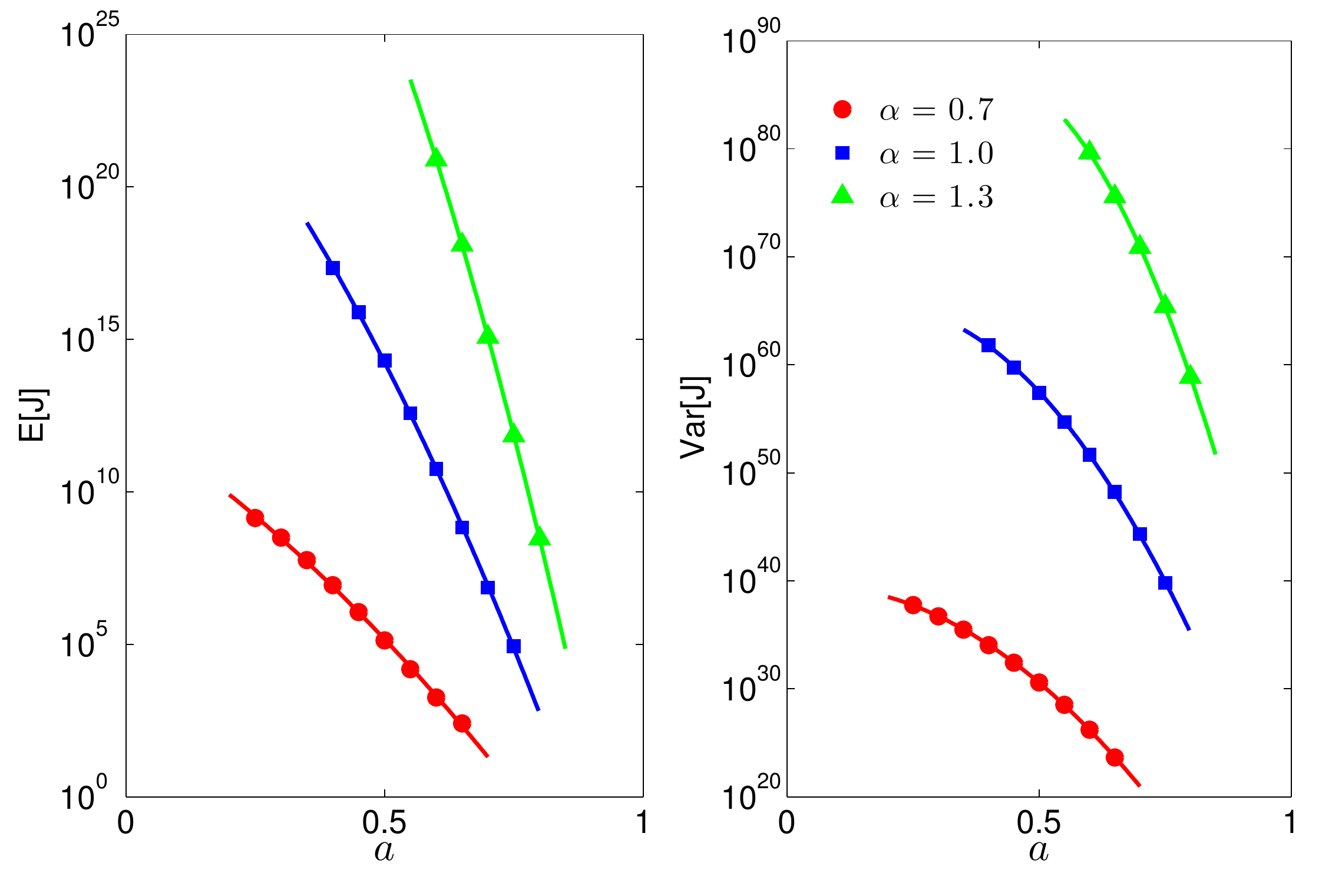}}
\caption{(Color on line) Average value of the coupling strength E[J] and of its variance Var[J], versus $a$ for several choices of parameters, depicted in different colors. We considered  $\alpha=0.7$ ($\circ$), $\alpha=1.0$ ($\square$) and $\alpha=1.3$ ($\triangle$) and $L=120$ (see the legend). Symbols refer to data obtained from exact numerical calculations, while curves are drawn according to the approximation~(\ref{eq:J_scaling}).
}
\label{fig:J_vs_a_L120}
\end{figure}

To obtain the mean coupling probability one should average Eq. (\ref{eq:p(J)}), so $P(\chi; \rho_i, \rho_j)$ of Eq. (\ref{eqn:prob_accoppiamento}), over the binomial distribution of $\rho_i$ and $\rho_j$ in Eq. (\ref{eqn:prob_ro}). To give an analytical estimate of this quantity, as in section \ref{sec:scal}, one can use a mean-field like approach, namely Eq. (\ref{pMF(chi)}).
Therefore, we can write the expressions:
\begin{eqnarray}
\left\langle \chi\right\rangle_{MF} = \sum_{\chi=0}^{(1-a)L} \chi \, P_{MF}(\chi) = \frac{1-a^2}{2}L,\\
\left\langle J\right\rangle_{MF} = \sum_{\chi=\lceil \frac{L}{\alpha+1}\rceil}^{(1-a)L}  P_{MF}(\chi) J(\chi),
\end{eqnarray}
and, as long as $\left\langle \chi\right\rangle_{MF}$ belongs to $[a^2 , \frac{\alpha+1}{\alpha-1}]$ we can approximate
\bea\label{eq:J_scaling}
\left\langle J\right\rangle_{MF} \simeq J(\left\langle \chi\right\rangle_{MF})  &=&
\mbox{exp}\left\{[(\alpha+1)\frac{1-a^2}{2}-1]L\right\} =
\nonumber\\
\phantom{\left\langle J_{MF}\right\rangle} &=&
N^{[(\alpha+1)\frac{1-a^2}{2}-1]\gamma}.
\eea
Hence, $\left\langle J\right\rangle_{MF}$ is expected to scale as a power of the system size and exponentially with $a^2$.
These results have been successfully checked by numerical simulations (see Figs. \ref{fig:J_vs_L} and \ref{fig:J_vs_a_L120}).

\subsection{Weighted connectivity}\label{sec:WC}

We now extend the bare degree $z_i=\sum_{j} A_{ij}$ to a \emph{weighted} degree referred to as $w_i$ and defined as
\begin{equation}
w_i = \sum_{j} J_{ij}.
\end{equation}
This quantity has a strong thermodynamic meaning, in fact, recalling the Hamiltonian of Eq. (\ref{eq:H_1}), and assuming, for the sake of simplicity, the zero noise limit so that all lymphocytes in the neighborhood of $i$ are quiescent, the local field acting on $i$ is just $\varphi_i = -\sum_{j=1}^N J_{ij} m_j = w_i$. This remark allows to establish a correlation between weighted degree of node $i$ and role of the $i$-th clone, i.e. either self or non-self addressed \cite{BA}.

\begin{figure*}[!] \label{fig:P_w}
\resizebox{1.2\columnwidth}{!}{\includegraphics{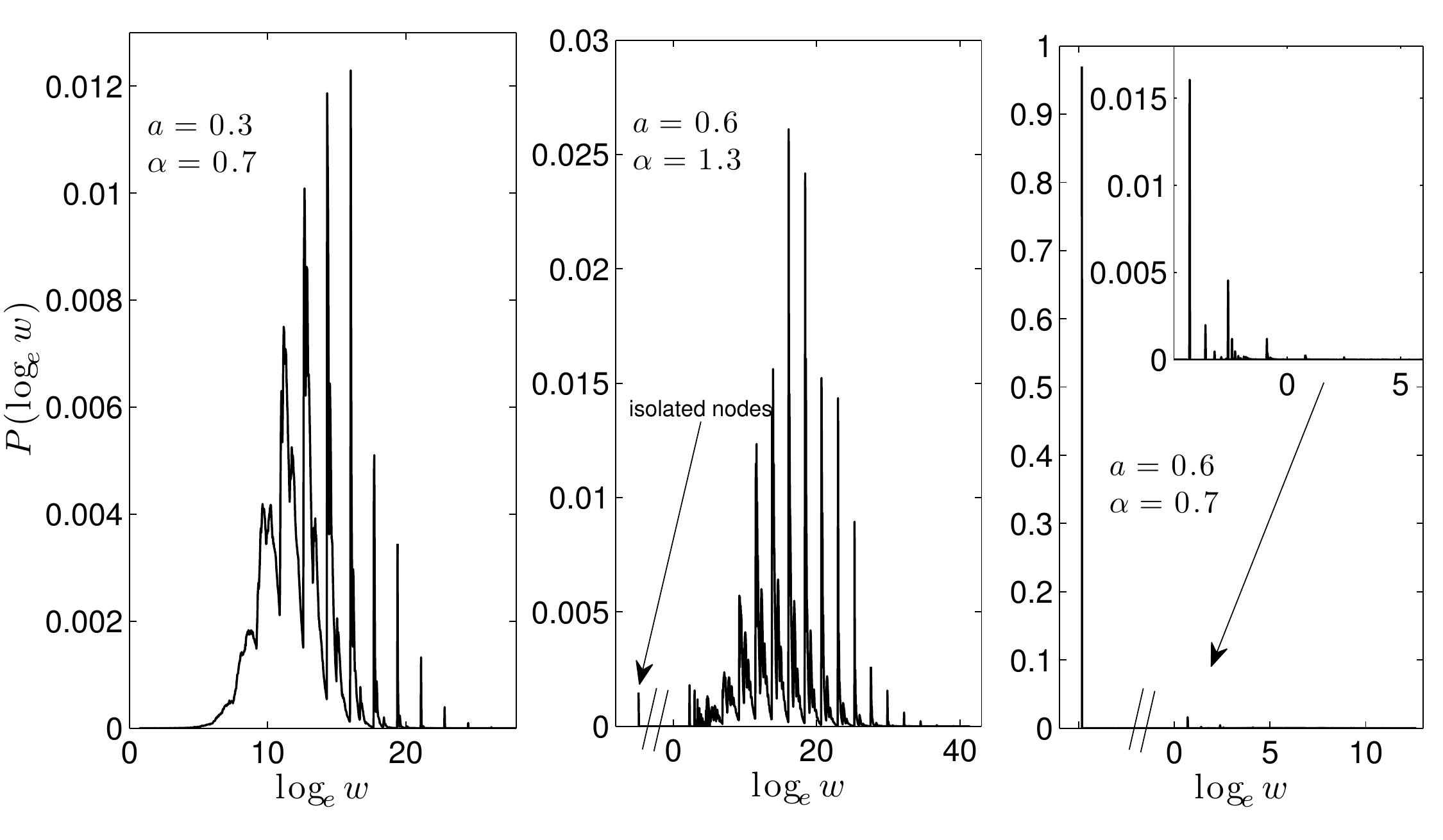}}
\label{fig:pw_a03_alpha07_N20000_L69_pro100}
\label{subfig:pw}
\caption{Semilogarithmic plot for the distribution $P(w;a,\alpha,L)$ obtained by averaging 100 systems made of $N=2 \times 10^5$ nodes with $\gamma=7$; results shown were averaged over $10^3$ different realizations. We considered several values of $a$ and $\alpha$, corresponding to either over- or under- percolated regimes.}
\end{figure*}

\begin{figure*}[!]
\resizebox{1.2\columnwidth}{!}{\includegraphics{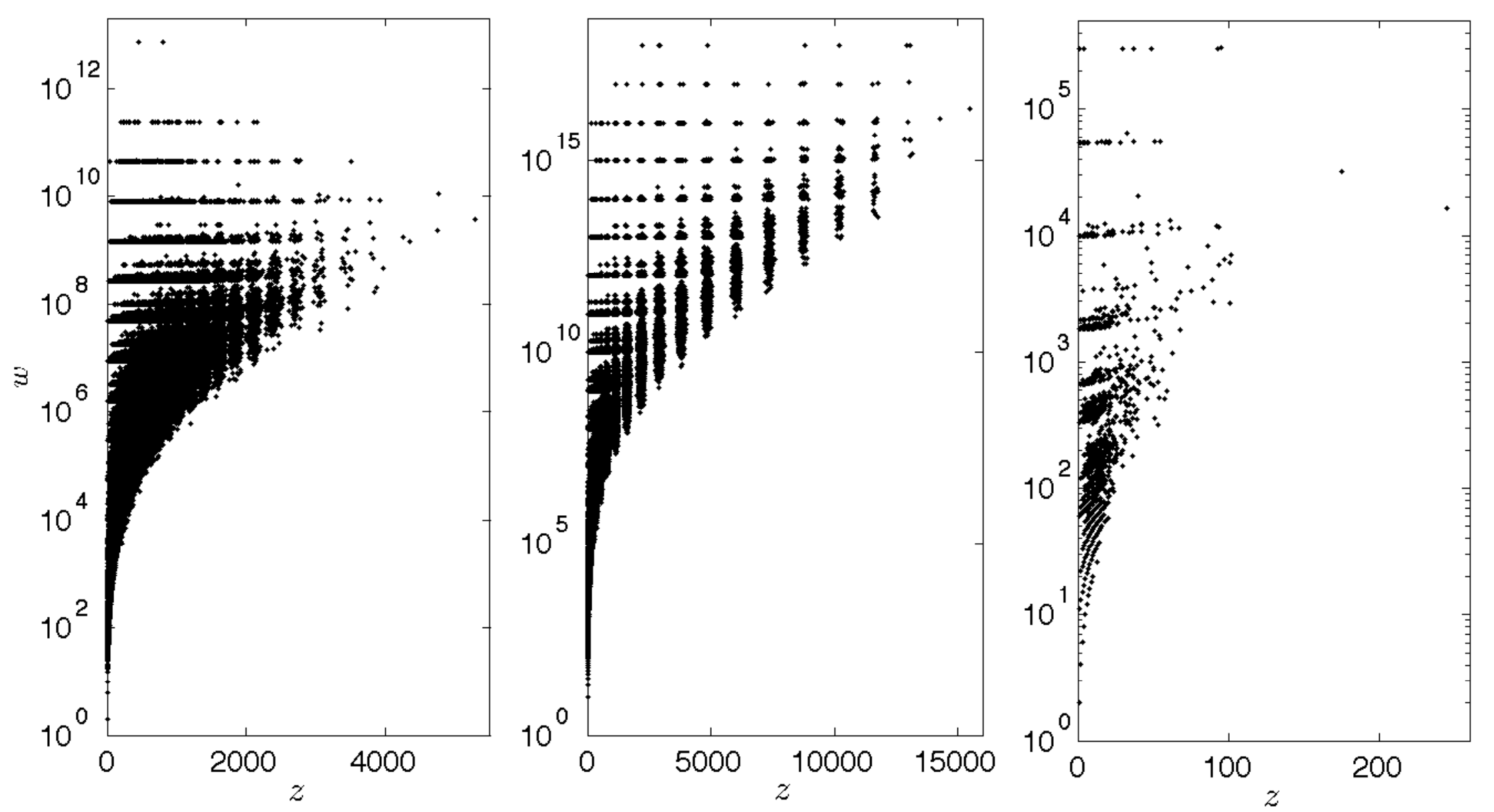}}
\label{fig:scatter_a06_alpha07_N20000_L69_pro100}
\label{subfig:scatter}
\caption{Each point shown corresponds to a node of the idiotypic network and its coordinates correspond to its weighted degree $w$ and its bare degree $z$, respectively. The 100 system simulated is made up of $N=2 \times 10^5$ nodes and we fixed $\gamma=7$. We considered the same values of $a$ and $\alpha$ used in Fig.~$7$. Notice that nodes displaying a different degree (and therefore a different $\rho$), may display the same weighted degree. The reshuffling among weighted and non-weigheted peaks is evident.}
\end{figure*}

In general, $w_i$ depends on the number of nearest-neighbors $z_{i}$ and on the coupling with each of them. So we define the weighted degree probability as follows:

\bea\label{p(w;rhoi,zi)}
&&\!\!\!\!\!\!\!\!\! P(w_{i}| \rho_{i}, z_{i}; a, \alpha, L) =
\nonumber\\
&&\!\!\!\!\!\!\!\!\! \sum_{J_{i1}....J_{i z_{i}}} P(J_{i 1}| \rho_i) \ldots P(J_{i z_{i} }| \rho_i)\, \delta(w_{i} - \sum_{j=1}^{z_{i}} J_{ij}) \,.
\eea

Now, averaging over all the possible numbers of nearest-neighbor,
\begin{equation}\label{eq:p(w;rhoi)}
P(w_{i}|\rho_{i}; a, \alpha, L) = \sum_{z_{i}} P(z_{i}|\rho_{i}; a, \alpha, L) P(w_{i}| \rho_{i} z_{i}; a, \alpha, L)
\end{equation}
and averaging over nodes we obtain the mean degree distribution
\beq\label{eq:p(w)}
P(w;a,\alpha,L) = \sum_{\rho_i} P(w_{i}|\rho_{i}; a, \alpha, L) P(\rho_i; a,L).
\eeq
This is the theoretical description of the distributions in Fig.~$7$ which were obtained by numerical simulations of the network. The weighted degree distribution displays a fine structure similarly to $P_{\textrm{degree}}(z)$ (see Sec. (\ref{sec:multimodal})), and even a ``hyperfine'' structure.
This is ultimately due to the fact that Eq. (\ref{p(w;rhoi,zi)}), being a sum of terms that can be localized in the $w$ range, can, by itself, display a multimodal distribution. When summing over $\rho_i$ in Eq. (\ref{eq:p(w)}), several multimodal distributions are superposed, giving rise to the complicated structure shown in Fig.~$8$. Otherwise stated, $\rho_i$ may univocally determine a range for the degree $z_i$ [leading to a multimodal $P_{\textrm{degree}}(z)$], but $z_i$, in turn, does not univocally determine a range for $w_i$. The ``reshuffling'' between bare and weighted degrees can be seen in the scatter plots in Fig.~$8$ and it is mirrored by the non-trivial structure of $P(w)$. As a result, nodes that are lazier in reacting to antigenic stimulation are not necessarily those with a large number of neighbors, but, rather, those with a large weighted degree; the two subsets cannot be trivially mapped into each other.

It is worth emphasizing that, as the weighted degree is a sum of exponential factors, the support of $P(w)$ covers several orders of magnitude, in both the connected and the disconnected regimes. This is consistent with the co-existence of agents highly (poorly) susceptible with respect to external stimuli, i.e. nodes with small (large) $w$. This difference has been attributed to the self-addressed or non-self-addressed attitude of lymphocytes \cite{stewart2,stewart3}, and, interestingly, it also survives in the underpercolated regime.

\begin{figure}[!]
\centering
\resizebox{1\columnwidth}{!}{\includegraphics{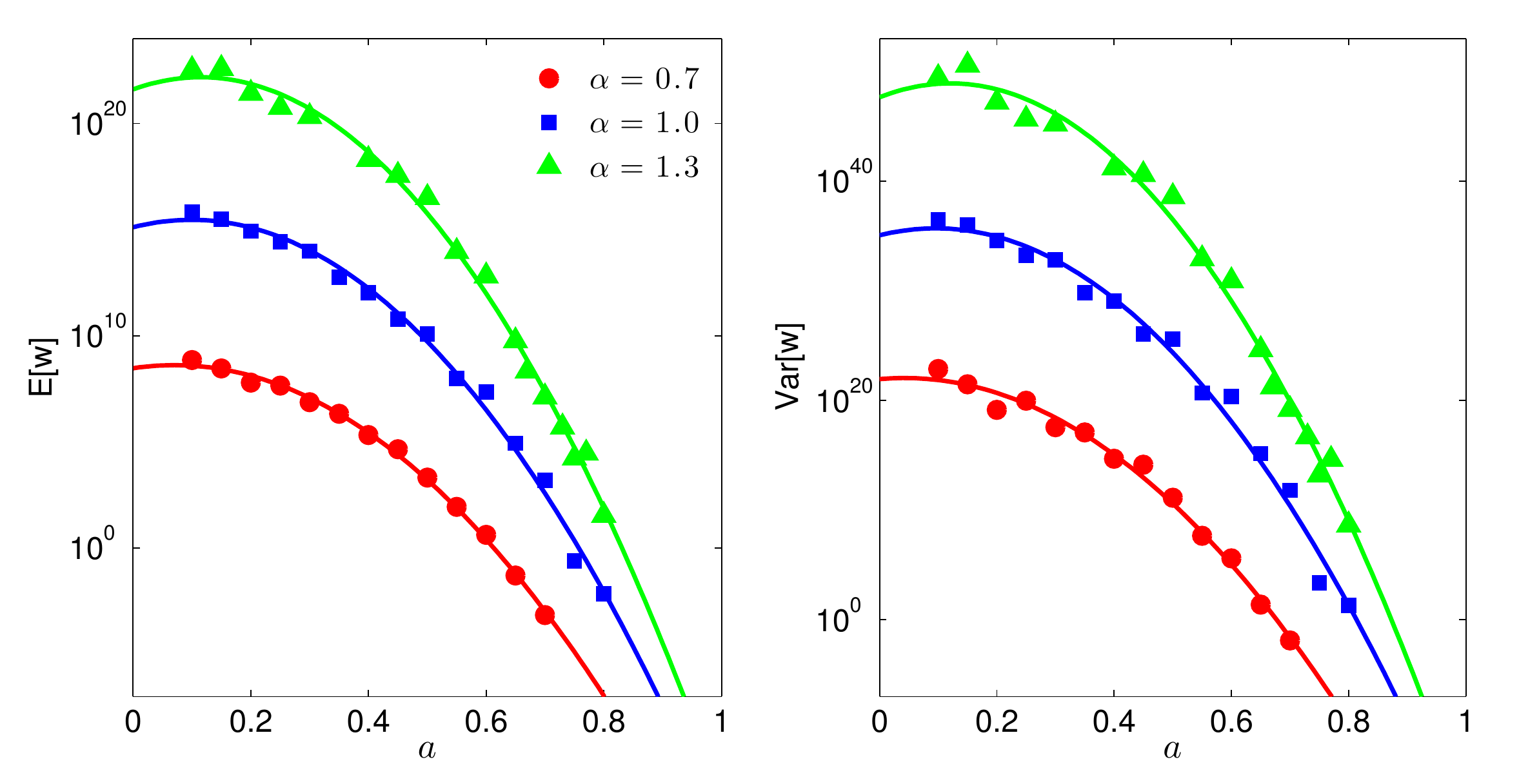}}
\caption{(Color on line) Average weighted degree $E[w]$ and its variance $Var[w]$ versus $a$ for a system of $N=10^4$ sites, $\gamma=7$ and $\alpha=0.7$ ($\circ$), $\alpha=1.0$ ($\square$), and $\alpha=1.3$ ($\triangle$). Symbols represent data from numerical simulations, while curves are the best fit obtained from Eq.~(\ref{eq:w_vs_J}).}
\label{fig:w_vs_a_alpha07_N10000_L64}
\end{figure}

In particular, we can introduce the relations
\bea\label{eq:w_vs_J}
\left\langle w \right\rangle &\sim& N \left\langle J \right\rangle,
\nonumber\\
\mbox{Var}\left[w \right] &\sim& N\, \mbox{Var}[J],
\eea
which hold as long as the couplings insisting on the same nodes can be approximated as independent. The expressions in Eq.~(\ref{eq:w_vs_J}) have been used to fit the numerical data in Fig.~(\ref{fig:w_vs_a_alpha07_N10000_L64}).

Finally, we stress that the support of $P(w)$ remains spread over several order of magnitude also in the region of the parameter space where the network is underpercolated (see Sec. \ref{sec:percola}): This suggests that the existence of an extensive \emph{Jerne-like} network may not be strictly necessary for \emph{self/non-self} discrimination in a systemic way.

\section{Bias and specificity} \label{sec:percola}

\begin{figure*}[!]
\centering
\resizebox{0.5\columnwidth}{!}{\includegraphics{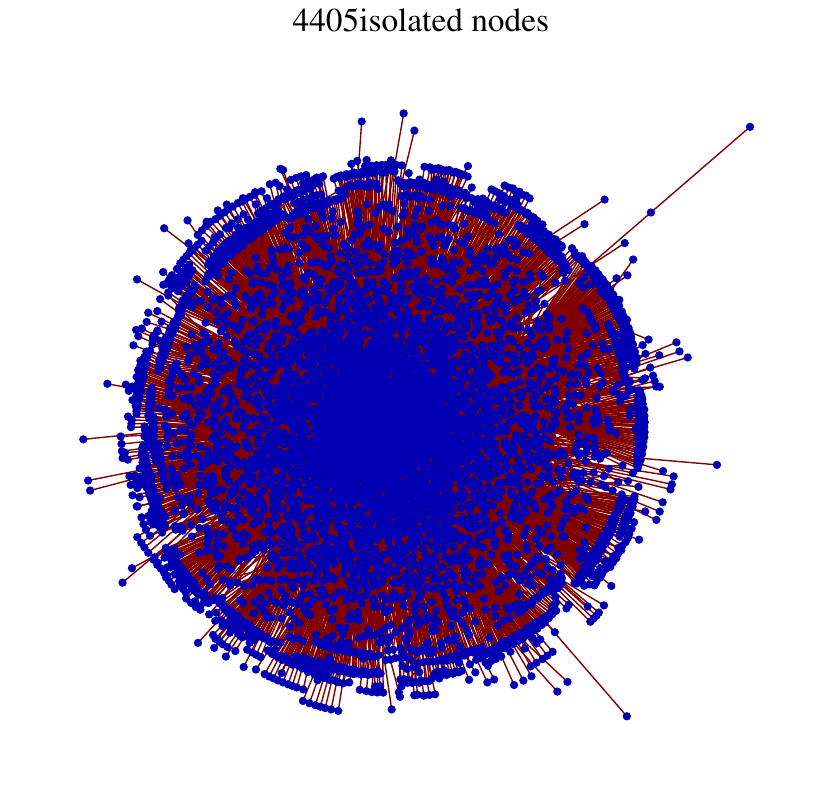}}
\resizebox{0.5\columnwidth}{!}{\includegraphics{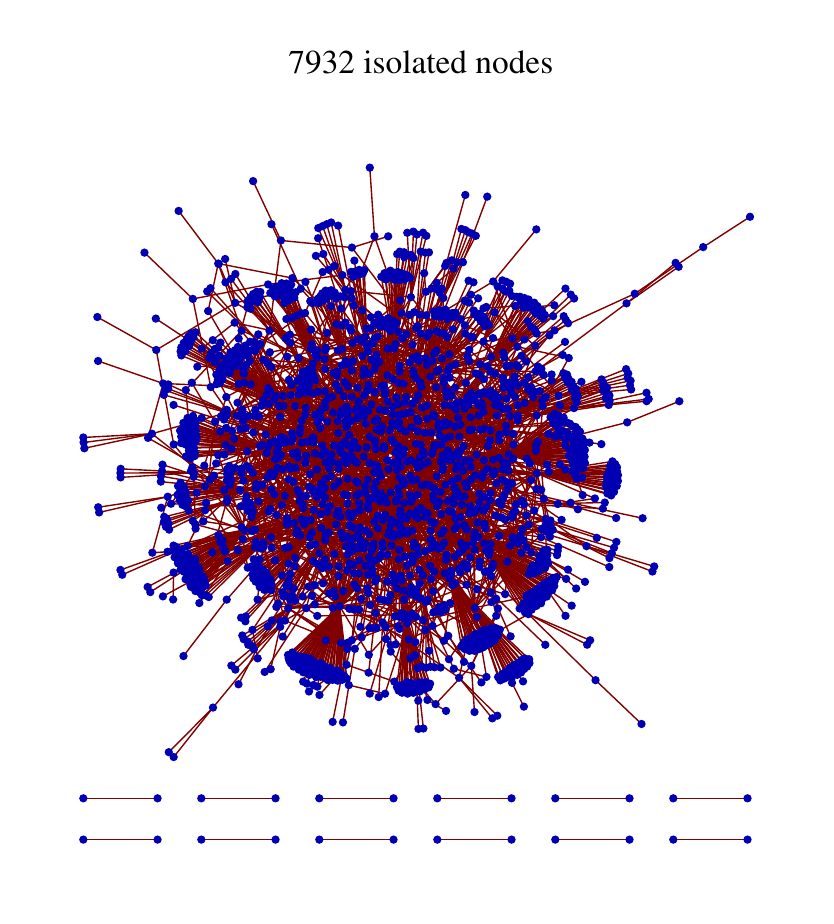}}
\resizebox{0.5\columnwidth}{!}{\includegraphics{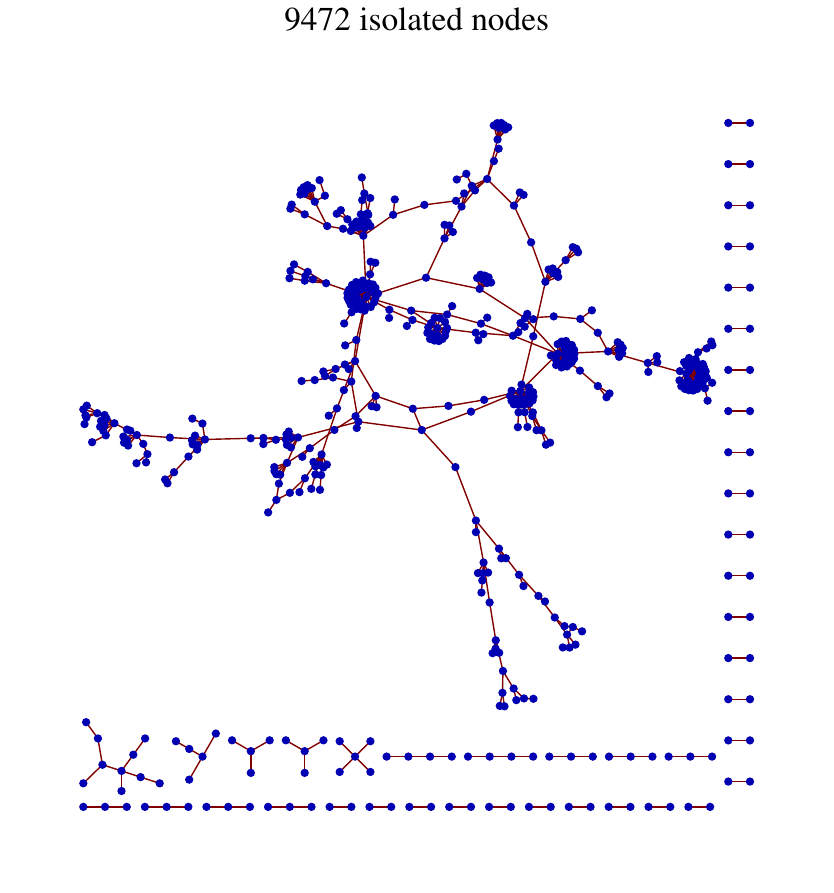}}
\resizebox{0.5\columnwidth}{!}{\includegraphics{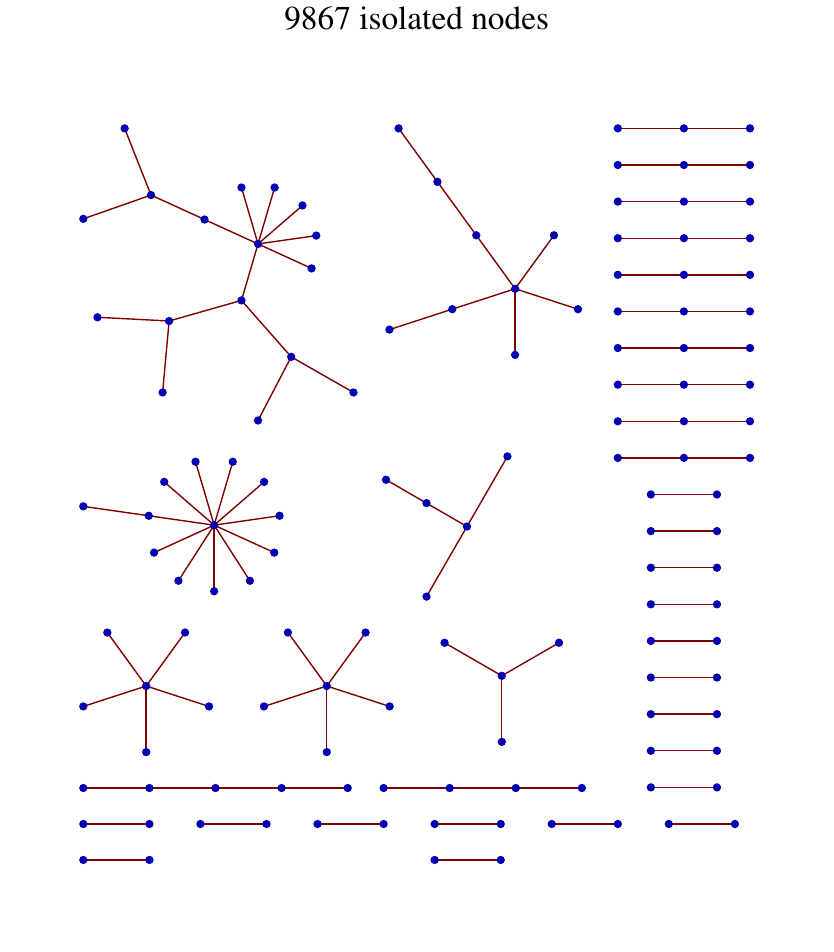}}
\caption{(Color on line) Realization of idiotypic networks made up of $N=10000$ nodes with $\gamma=3$ and $\alpha=0.7$. Different values of $a$ have been considered: from left to right $a=0.50, \,0.55,\,0.59,\,0.62$. Although these plots refer to one single realization, we have checked that the system displays robustness in this sense.}
\label{fig:Mathematica}
\end{figure*}

As explained before, an increase in the bias parameter $a$ corresponds to a progressive smoothing of the repertoire variability. A growth in $a$ also results in a dilution of the graph itself, eventually leading to percolation phenomena (see Fig.~(\ref{fig:Mathematica})).

In order to study this process, we compare its features to those pertaining to an ER random graph $\mathcal{G}_{\textrm{ER}}(N,p)$, where links are drawn independently with probability $p=\bar{z}/N$, in such a way that the average degree is the same for both networks.

Therefore, the evolution of $\mathcal{G}(N,p)$ as $p$ ranges from $0$ to $1$, eventually leading to a percolation transition (see, e.g., \cite{ER,bela}) is compared to the evolution of our graph $\mathcal{G}(a,L,\alpha)$ as $a$ ranges from $0$ to $1$.

\subsection{Giant component and distribution of cluster size}
In order to evaluate the impact of removing ties, we measure
the relative size of the largest connected component $S$ as a
function of the fraction of links left $f$ .

In general, as $a$ ranges from $0$ to $1$, $S$ gets continuously smaller: nodes with large $\rho$ are those more likely to remain
isolated or to form small clusters. Differently from the ER case, beyond the giant component, clusters typically display a small size.
As evidenced in Ref. \cite{ACG}, these features give rise to a rather gentle percolation transition.

In order to clarify this point, we focus on the evolution of the
internal organization of clusters by measuring the distribution
$N(a,s)$, representing the number of clusters of size $s$ present when the correlation parameter is $a$.

\begin{figure}[!]
\centering
\resizebox{0.9\columnwidth}{!}{\includegraphics{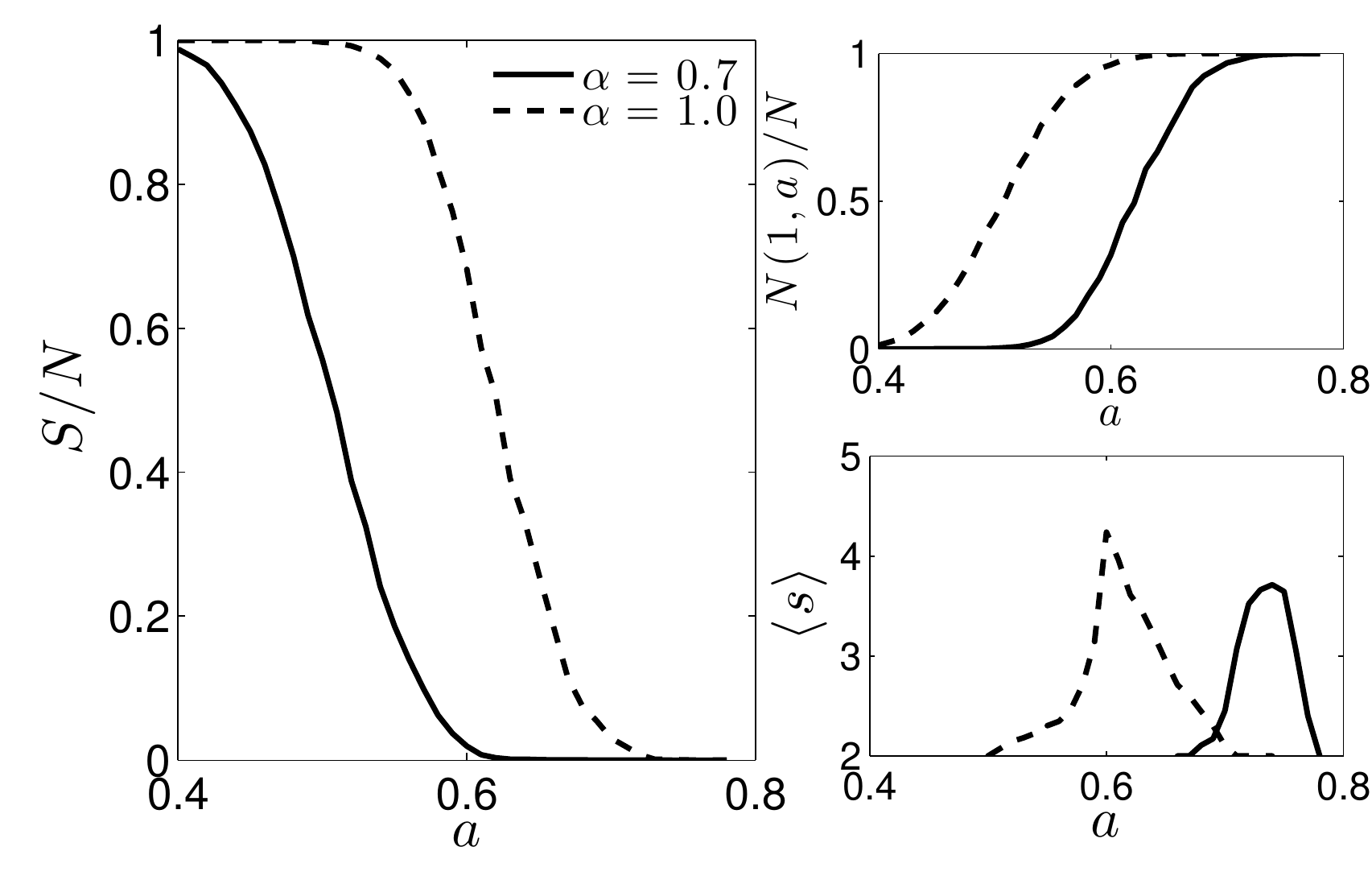}}
\caption{ These plots describe the evolution of the network as $a$ is tuned while we fix the system size $N=10000$, $\gamma=7$, and $\alpha=0.7$ (solid line) and $\alpha=1$ (dashed line).
Left panel: Average size of the giant component $S/N$. Right panels: relative number of isolated nodes $N(1,a)/N$ (upper panel) and average size of the non-giants components excluding isolated nodes (lower panel).}
\label{fig:bar3}
\end{figure}

As shown in Fig.~(\ref{fig:bar3}), as $a$ grows, the typical cluster size shrinks from $N$ (a unique giant component) to $1$ (there are only isolated nodes). For instance, at $\alpha=0.7$, when $a=0.46$  there are a few isolated nodes and a giant component whose typical (normalized) size is around $0.8$. On the other hand, when $a=0.60$ most nodes are isolated, $S \approx 0.39$ and the typical size of the remaining nodes is around $4.2$. Beyond isolated nodes and the giant component, the statistics of cluster size is rather uniform, suggesting that minor disconnected clusters display small sizes, i.e., $s < 10$.


\subsection{Emergence of small components}
\begin{figure}[!]
\begin{center}
\resizebox{1.0\columnwidth}{!}{\includegraphics{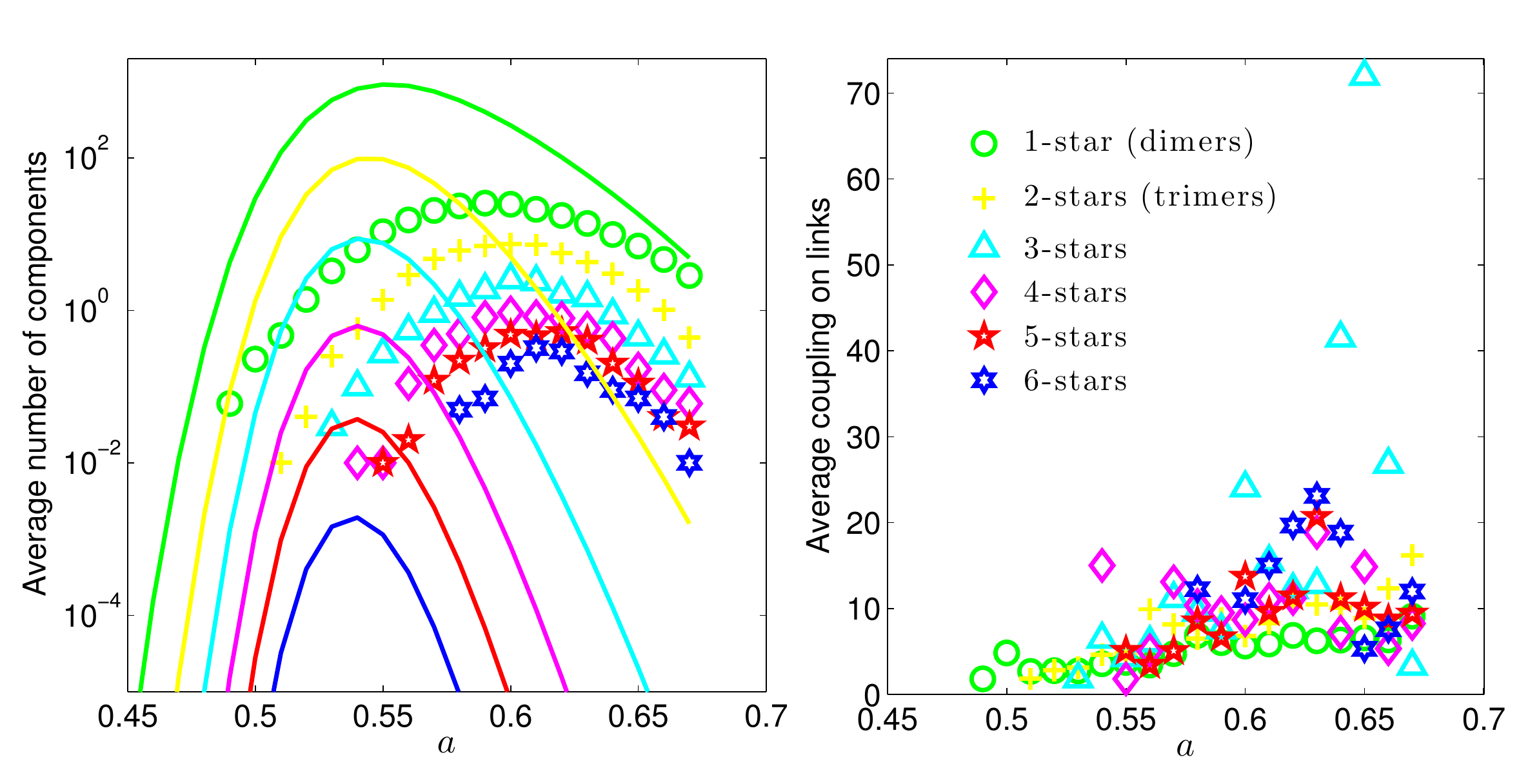}}
\caption{Left: Number of $k$-stars (a dimer is a 1-star and a trimer is a 2-star) averaged over 100 numerical realizations of $N$=10000 and $\alpha$=0.7 idiotypic networks, while $a$ (circles) is variated. Solid lines represent the ER prediction for the same quantity at the relative link probability. Right: Mean coupling value within a star.}
\label{fig:animali_N10000_alpha07}
\end{center}
\end{figure}

The difference between our idiotypic network and a random graph is emphasized by the analysis of the emergence of motifs around the the percolation threshold. We study the number of isolated $k$-loops, namely, unicycles of length $k$, and isolated $k$-stars, namely, subgraphs made of a central node connected to $k$ nodes with unitary degree. In an ER graph these quantities follow
\begin{eqnarray}
\!\!\!\!\left\langle N_{\textrm{$k$-star}}\right\rangle =  p^k (1-p)^{(k+1)N-3k-1} \frac{\prod_{i=0}^k(N-i+1)}{(k+1)!},\\
\!\!\!\!\left\langle N_{\textrm{$k$-loop}}\right\rangle = p^k (1-p)^{k(N-k)} \frac{\prod_{i=0}^k(N-i+1)}{k!},
\end{eqnarray}
which are drawn for comparison in Fig. \ref{fig:animali_N10000_alpha07}. In the idiotypic network the above-mentioned components are present in a smaller range of $a$, and the number of all stars, except the dimers ($1$-star), is eventually larger than ER predictions. This is due to the bias characterizing strings, as it makes the percolation transition smoother - in such a way that components not belonging to the giant component hardly develop - and induces inhomogeneity among nodes - in such a way that those associated with a value of $\rho$ much lower than its average value are likely to act as (local) hubs. This picture is confirmed by the fact that $k$-stars with large $k$ have a higher mean coupling (see Fig. \ref{fig:animali_N10000_alpha07}, right).

Moreover, in our simulations we never find isolated triangles (3-loop) or quadrangles (4-loop), which are instead present in random graphs.

\section{Conclusions}\label{sec:concl}
In this work we studied a class of weighted random networks aimed to describe the mutual interactions between cells inside the B-branch of the immune system.
As in previous works \cite{immuno,physa}, the Burnet clonal expansion theory \cite{a6,a48}  appears to be the standard one-body response of the system described by the Hamiltonian of Eq.~(\ref{eq:H_1}), while the Jerne theory is ruled by the two-body term, whose coupling encodes the Varela prescription.

The main novelty introduced is the existence of a biased repertoire: The parameter tuning the "mean field similarity" among the epitopes (the entries in the bit-strings encoding for antibodies) is a scalar $a$ ranging from $0$ (completely random) to $1$ (completely deterministic).
From a biological viewpoint, the extension to a value $a \neq 0$ is a step toard more realistic descriptions, since antibodies are not completely random objects \cite{kardar} and since during  ontogenesis in bone marrow, a part of the strongly self-reacting repertoire is killed \cite{a11,franz}. General findings \cite{ciarle2,ciarle1} and recent investigations (achieved in the study of the zebra fish repertoire \cite{bialek1}\cite{bialek2}) show that the ensemble of the genetic  heritage  dedicated to antibody formation is used in a highly inhomogeneous (Zipf-like) way,  suggesting that some epitopes can be highly over-expressed compared to others. Although this would imply the use of an $a_i^{\mu}$ (with the price of a highly untractable mathematics), we skipped the fine dependence in favor of a simpler mean field choice $a_i^{\mu}= a \neq 0$ for all the epitopes \footnote{Clearly biasing the repertoire, both the networks gained trough $a_i^{\mu}$ or $a$ share a reduction of the connectivity because the over-expression of a particular epitope increases the probability of finding it in two randomly chosen strings and consequently lowers their binding affinity.}.
 This introduction of bias indeed gives rise to striking, at least qualitatively, effects: In a broad range of $a \neq 0$, the degree distribution displays a multi-modular structure which mirrors the history of the system itself. Such fringes in the (weighted) degree distribution were experimentally noted in the pioneering investigations by Varela and coworkers and an immunological interpretation was provided in Refs. \cite{stewart1,stewart2,stewart3}. Although based on synthetic data, our analysis extends these findings in some details, suggesting the existence of a hyperfine structure, richer than the one obtained by sampling over the experimentally available subset of the repertoire. Furthermore, the reshuffling among modes of the network raises the question of which should be the physical observable to characterize the reaction attitude of nodes:
the bare degree (which is actually the standard one checked in experiments) or, rather, the weighted connectivitys which we claim to be the most relevant; We hope to report further investigation on this point soon.
Here, we just stress that, while the average degree varies over a linear scale, the weighted degree displays a distribution with a log-normal like envelope, hence spanning several order of magnitude. This spread is robust as it holds also also for regimes of high dilution (corresponding to large $a$): Interestingly, this implies that mechanisms such as self/non-self discrimination may work despite a truely over-percolated network of B-cells. 

A progressive increase in $a$ eventually leads to an under-percolated network, where nodes not belonging to the giant component typically form small-size clusters, which mirrors the cascades of anti-antibodies commonly seen in experiments (see e.g. \cite{cazenave}).  Further, the local topology displays squares as motifs and triangles as non-motifs, as expected from a complementary-based network, which is in agreement with experiments on idiotypic networks too.

Moreover, we stress that the general picture in which the network starts with a high connectivity, then specializing its responses lets the coordination number decrease (and $a$ grows in our model)  describes here a "thermodynamical growth process" as, once the network reaches the percolation threshold, its entropy  remains positive even at a zero noise level: allowing a dynamics on $a$, that is slower with regard to the $\sigma$, the evolution "naturally" shifts $a$ to a nonzero value to maximize the entropy of the system \footnote{Below the percolation threshold, a positive entropy even in the zero noise limit is clearly not a violation of Nernst theorem because the system trivially factorizes in independent subsystems.}.

In future investigations, several experimental-based refinements and improvements are in order to achieve better predictive power. In particular, a change in the idiotope/paratope distributions toward mathematically challenging but biological more plausible choices \cite{bialek2} should be analyzed, and a clear relation among the mature repertoire and the bias received during ontogenesis \cite{cancro} should be determined too.

\bigskip

\section*{Acknowledgments}
This work was supported by FIRB Grant No. RBFR08EKEV.
\newline
Sapienza Universit\`a di Roma, Istituto Nazionale di Fisica Nucleare, and Gruppo Nazionale per la Fisica Matematica are also acknowledged.

\end{document}